\documentclass[fleqn,usenatbib]{mnras}

\usepackage{newtxtext,newtxmath}

\usepackage[T1]{fontenc}

\DeclareRobustCommand{\VAN}[3]{#2}
\let\VANthebibliography\thebibliography
\def\thebibliography{\DeclareRobustCommand{\VAN}[3]{##3}\VANthebibliography}

\usepackage{graphicx}	
\usepackage{amsmath}	
\usepackage{color}
\usepackage{threeparttable}

\title[Multi-path depolarization]{Faraday depolarization and induced circular polarization by multi-path propagation with application to FRBs}
\author[Beniamini, Kumar \& Narayan]{
	Paz Beniamini$^{1,2,3}$\thanks{pazb@openu.ac.il}, Pawan Kumar$^4$\thanks{pk@astro.as.utexas.edu}, Ramesh Narayan$^{5,6}$\thanks{rnarayan@cfa.harvard.edu}\\
	$^{1}$Department of Natural Sciences, Open University of Israel, 1 University Road, 43107 Ra'anana, Israel\\
	$^2$Astrophysics Research Center of the Open University (ARCO), The Open University of Israel, P.O Box 808, Ra’anana 43537, Israel\\
	$^3$Theoretical Astrophysics, Walter Burke Institute for Theoretical Physics, Mail Code
	350-17, Caltech, Pasadena, CA 91125, USA\\
	$^4$Department of Astronomy, University of Texas at Austin, Austin, TX 78712, USA\\
	$^5$Center for Astrophysics | Harvard \& Smithsonian, 60 Garden Street, Cambridge, MA 02138, USA\\
	$^6$Black Hole Initiative at Harvard University, 20 Garden Street, Cambridge, MA 02138, USA
}

\begin{document}
	\label{firstpage}
	\pagerange{\pageref{firstpage}--\pageref{lastpage}}
	\maketitle
	
	\begin{abstract}
		We describe how the observed polarization properties of an astronomical object are related to its intrinsic polarization properties and the finite temporal and spectral resolutions of the observing device. Moreover, we discuss the effect that a scattering screen, with non-zero magnetic field, between the source and observer has on the observed polarization properties. We show that the polarization properties are determined by the ratio of observing bandwidth and coherence bandwidth of the scattering screen and the ratio of temporal resolution of the instrument and the variability time of screen, as long as the length over which the Faraday rotation induced by the screen changes by $\sim\pi$ is smaller than the size of the screen visible to the observer. We describe the conditions under which a source that is 100\% linearly polarized intrinsically might be observed as partially depolarized, and how the source's temporal variability can be distinguished from the temporal variability induced by the scattering screen. In general, linearly polarized waves passing through a magnetized scattering screen can develop a significant circular polarization. We apply the work to the observed polarization properties of a few fast radio bursts (FRBs), and outline potential applications to pulsars. 
	\end{abstract}
	
	
	
	
	\section{Introduction}
	\label{sec:Intro}
	Fast Radio Bursts (FRBs) are powerful radio signals of usually a few ms duration (as short as 30 $\mu$s), which have been detected between $\sim$100 MHz and 8 GHz, with a typical flux of $\sim 1$ Jansky (10$^{-23}$ erg s$^{-1}$ cm$^{-2}$ Hz$^{-1}$) {\it e.g.} \cite{Lorimer+07,Thornton+13,Spitler2014,Petroff2016,Bannister2017,Law+17,Chatterjee+17,Marcote2017,Tendulkar+17,Gajjar2018,Michilli+18,Shannon2018,Oslowski2019,Kocz2019,Bannister+19,CHIME2019,CHIME2019b,Ravi2019,Ravi2019b,CHIME_1st_cat}.
	At the time of writing, 794 sources and 24 repeating sources consisting of more than a few hundred pulses altogether \footnote{\url{https://www.wis-tns.org/}} \citep{Spitler+16,CHIME2019,Kumar2019,CHIME_repeaters,CHIME_1st_cat}. The rate of occurrence of FRBs is $820 \pm 60\mbox{ (stat.)}^{+200}_{-220}\mbox{ (sys.)}$ at 600 Mhz \citep{CHIME_1st_cat}.

	Several FRBs are found to be 100\% linearly polarized (FRBs 20121102A, 20171209A, 20190604A, 20190711A), and quite a few have linear polarization between 30 and 90\% (FRBs 20190102C, 20190608B, 20180311A, 20180714A, 20181112A, 20160102A, 20151230A \& 20150807A), e.g. \cite{Michilli+18,Oslowski2019,CHIME_repeaters,Cho2020,Luo2020}. A few FRBs (20140514A, 20180309A, 20190611B, 20201124A) are found to have non-zero circular polarization between 20 and 60\%, e.g. \cite{Petroff+2014,Oslowski2019,Day+2020,Hilmarsson2021}. The diversity of the observed polarization properties is likely to hold important clues regarding the FRB emission mechanism and propagation effects in the magneto-ionic medium between the source and us.
	
	Previous studies have discussed a physical effect that could convert an initially linearly polarized wave to a partially circularly polarized wave due to its propagation through a magneto-ionic medium, which is referred to as "Generalized Faraday rotation" or "Faraday conversion" \citep{Cohen1960,ZZ1964,KM1998,VR2019,GL2019}. The generalized Faraday rotation typically produces a small amount of circular polarization. The effect relies either on (i) 
	propagation through a strongly magnetized region with magnetic field reversals (the expected circular polarization in this scenario is $\Pi_{\rm cir}\sim 3\times10^{-8}\nu_{\rm Ghz}^{-2}\mbox{RM}_3B_{\rm \mu G}$ where RM$_3$ is the rotation measure in units of $10^3\mbox{rad m}^{-2}$ and $B_{\rm \mu G}$ is the field strength in $\mu G$.) or (ii) departure of the EM wave dispersion relation from that of a cold magnetized plasma. In the latter case, one requires again a large magnetic field as well as a plasma consisting of mildly relativistic electrons in order to produce any significant amount of circular polarization.

	This work focuses on a physically distinct mechanism, viz. multi-path propagation of a wave through a magnetized scintillating screen. With standard parameters expected for ISM-like screens (and in particular much lower magnetic field values than those mentioned above), multi-path propagation can significantly change the observed polarization and potentially induce a large degree of circular polarization. After presenting a self-contained description of how polarization properties are modified due to propagation of radio waves through a magnetized scattering screen, we discuss whether FRBs with less than 100\% linear polarization, and the ones with circular polarization, might be the result of the propagation effects. The physics described here is also applicable to pulsars and other astrophysical systems. A specific application to other systems will be the topic of a future investigation.

	The layout of this paper is as follows. In \S \ref{sec:tempvar} \& \ref{sec:nu} we discuss depolarization due to the finite temporal and spectral resolutions of observations. In \S \ref{sec:Screen} we derive general results for depolarization resulting from propagation of EM waves through a magnetized, turbulent, scattering screen, and apply those results to FRB observations in \S \ref{obs}. The main results of this work are summarized in \S \ref{sec:discuss}.
	
	\section{Depolarization due to finite temporal resolution}
	\label{sec:tempvar}
	Consider a quasi-monochromatic electromagnetic wave,
	\begin{equation}
		\vec{E}=\hat{x}E_1+\hat{y}E_2 \quad ; \quad E_{\rm j}=A_{\rm j}(t)\exp{[i(\phi_{\rm j}(t) -\omega t)]} \quad \mbox{for } j=1,2
	\end{equation}
	such that its period is $T = 2\pi/\omega$, and the timescale for fluctuations in its amplitudes and phases is $t_{\rm var}\gg T$. A measurement of the observed polarization of the wave involves a detector with a finite integration timescale $t_{\rm res}\gg T$, and gives us Stokes parameters which are linear combinations of the terms
	\begin{equation}
		\label{eq:EiEj}
		\langle E_{\rm i}E_{\rm j}^*\rangle_t=\frac{1}{t_{\rm res}}\int_0^{t_{\rm res}} E_{\rm i}(t)E_{\rm j}^*(t)dt.
	\end{equation}
	where $i,j=1,2$ and $E_1,E_2$ are the complex components of the electric field along two perpendicular directions in the plane of the sky. Note that for monochromatic waves, the time dependence enters only through $\exp{[i\omega t]}$, and so the integrand in Eq. \ref{eq:EiEj} is time-independent for any $i,j$. The Stokes parameters for the quasi-monochromatic wave are
	\begin{eqnarray}
		\label{eq:Stokes}
		& I=\langle E_1E_1^*\rangle_t + \langle E_2E_2^*\rangle_t \quad ; \quad Q=\langle E_1E_1^*\rangle_t - \langle E_2E_2^*\rangle_t \nonumber \\
		& U=\langle E_1E_2^*\rangle_t + \langle E_2E_1^*\rangle_t \quad ; \quad V=\frac{1}{i}(\langle E_1E_2^*\rangle_t - \langle E_2E_1^*\rangle_t).
	\end{eqnarray}
	The linear, circular, and the overall degree of polarization are described by
	\begin{equation}
		\Pi_{\rm lin,t}=\frac{\sqrt{Q^2+U^2}}{I}, \quad \Pi_{\rm cir,t} = \frac{V}{I}, \quad \Pi_t=\frac{\sqrt{Q^2+U^2+V^2}}{I}\leq 1,
	\end{equation}
	where the index $t$ indicates that this equation captures temporal depolarization and where equality holds for a wave that is monochromatic over the integration period, such that $\langle E_{\rm i}E_{\rm i}^*\rangle_t \langle E_{\rm j}E_{\rm j}^*\rangle_t=\langle E_{\rm i}E_{\rm j}^*\rangle_t \langle E_{\rm j}E_{\rm i}^*\rangle_t$.
	We calculate the degree of polarization by means of a Monte Carlo simulation and analytic estimates for limiting cases detailed below. Given a timescale for fluctuations $t_{\rm var}$, we divide the integration time into sections of length $t_{\rm var}$ and randomly change the amplitudes / phases of the wave within each new section. In particular we focus on a situation in which the wave within each time section is 100\% linearly polarized, and between sections only the wave's polarization angle (PA), and not its amplitude, are changed.
	We consider two limiting cases. The ``no memory" case, in which 
	the PAs in each time section are randomly drawn from a uniform distribution in the range $[0,\Delta \chi]$.  This choice implies that the source has ``no memory", i.e. that the change in PA produced by the source at times differing by $\Delta t\gg t_{\rm var}$ is independent of $\Delta t$. An opposite extreme, is the ``random walk" limit in which the PA produced by the source is strongly correlated with the PA produced at the previous time step (i.e. $\Delta \chi$ relative to the previous time section is drawn from the same distribution). 
	
\begin{figure}
		\centering
		\includegraphics[width = 0.4\textwidth]{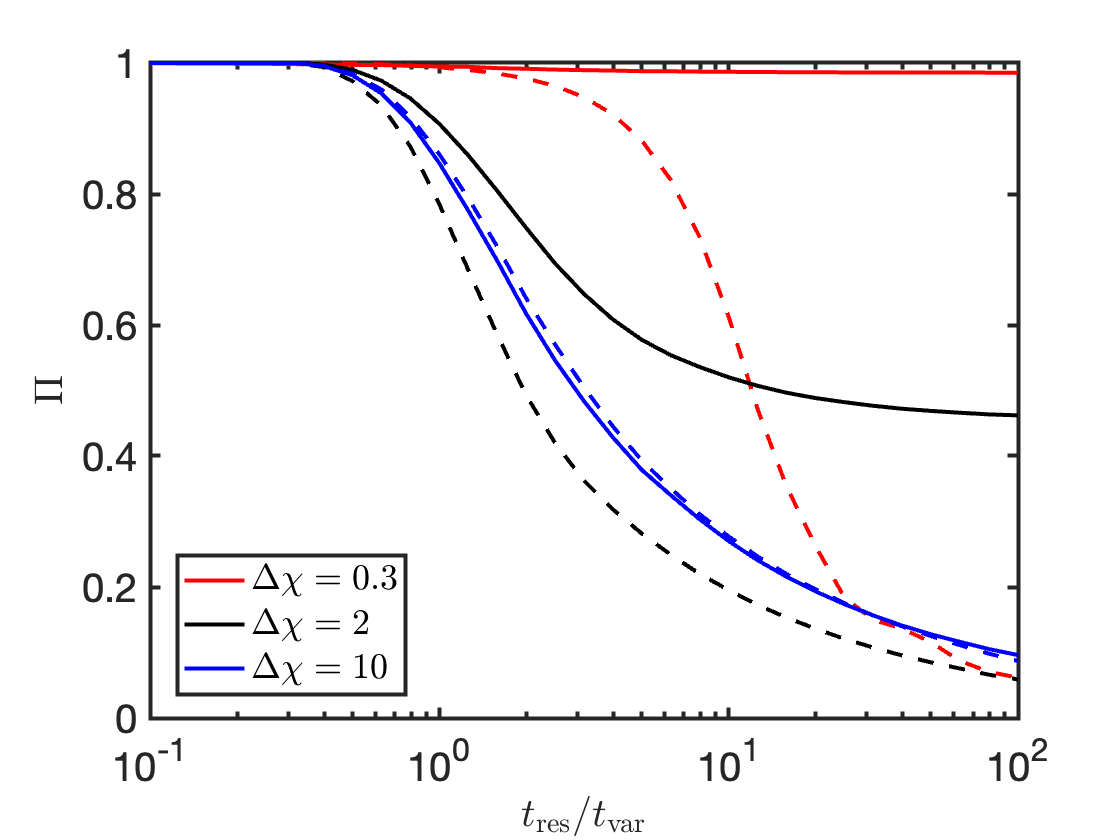}
		\caption{Degree of polarization for a linearly polarized wave with a PA that changes on a timescale $t_{\rm var}$ uniformly within the range $[0,\Delta \chi]$. The temporal resolution of the instrument is $t_{\rm res}$. Solid lines are calculated under the `no memory' assumption and dashed curves correspond to the `random walk' case.} In all cases the degree of circular polarization of the outgoing wave is zero.
		\label{fig:tempres}
	\end{figure}

	Regardless of the overall degree of polarization recorded by the detector, the measured degree of circular polarization in this scenario, is always zero ($V=0$).
	The asymptotic degree of polarization (i.e. for $t_{\rm res}\gg t_{\rm var}$) can be analytically understood in two simplifying cases. The first case we consider is that of large swings in the PA from one coherent segment to next, i.e. $\Delta \chi \gg 1$ (note that in this limit the ``no memory" and ``random walk" cases become indistinguishable).  The contributions to $Q,U$ from the $N_t=t_{\rm res}/t_{\rm var}$ independent segments add up in a random walk fashion, while the intensities, $I$, scale linearly with $N_t$. The result is
	\begin{equation}
		\label{eq:delchigg1}
		\Pi_t \approx \frac{\sqrt{Q^2 +U^2}}{ I }=\frac{\sqrt{N_t}}{N_t}=\sqrt{\frac{t_{\rm var}}{t_{\rm res}}} \quad \quad \mbox { for}\quad \Delta \chi \gg 1, t_{\rm res}\gg t_{\rm var}
	\end{equation}
	or 
	\begin{equation}
		\Pi_t=\left(\frac{t_{\rm res}}{t_{\rm var}}+1\right)^{-1/2} \quad \quad \mbox { for}\quad \Delta \chi \gg 1,\mbox{ general $N_t$}.
	\end{equation}
	Next, consider the ``no memory" case with $\Delta \chi\ll1$ and an incoming wave which is pure Q. For $t_{\rm res}>t_{\rm var}$, we can expand $Q,U$ of the outgoing wave in small $\Delta\chi$, and obtain $Q/I=\langle 1-2\chi^2 \rangle_{\chi}=1-\frac{2}{3}\Delta \chi^2$ (where the subscript $\chi$ denotes averaging over $\chi$ in the range [$0,\Delta\chi$], which, in the particular case of the setup described above, is equivalent to averaging over time) and $U/I=\langle 2\chi \rangle_{\chi} =\Delta \chi$, so that
	\begin{equation}
		\label{eq:delchill1}
		\Pi_t \approx \frac{\sqrt{Q^2 +U^2}}{I}\approx 1-\frac{1}{6} \Delta \chi^2 \quad \mbox { ``no memory", } \Delta \chi \ll 1, t_{\rm res}\gg t_{\rm var}.
	\end{equation}
	We conclude that under the ``no memory" assumption (in which the PAs are always confined to a finite range $[0,\Delta \chi]$) strong depolarization is only possible when $t_{\rm res}\gg t_{\rm var}$ and $\Delta \chi\gg 1$. In the ``random walk" case, depolarization occurs if $N_t^{1/2}\Delta \chi \gg 1$, even if $\Delta \chi<1$ (and in that limit $\Pi_{\rm t}\approx N_{\rm t}^{-1/2}$ as for $\Delta \chi>1$).
	The results of our Monte Carlo calculation are shown in figure \ref{fig:tempres}. The asymptotic limits described above are reproduced by the Monte Carlo calculation.

	\section{Depolarization due to finite spectral resolution}
	\label{sec:nu}
	\subsection{Stochastically varying wave properties}
	\label{sec:nuvar}
	The observed level of polarization of a wave depends also on the detector's bandwidth wrt the intrinsic spectral variability in a way that is similar to its dependence on the temporal variability relative to the temporal resolution of the detector.
	Consider a signal that is observed over a certain frequency range. Let us denote the spectral resolution of the detector by $\nu_{\rm res}$, and the minimum spectral bandwidth over which the wave's electric field direction changes by $\sim\pi$ to be $\nu_{\rm co}$. Both these frequencies are taken to be small relative to the central frequency of the signal $\nu_{\rm res}, \nu_{\rm co}\ll \nu$. To focus on the case of depolarization due to finite spectral resolution, we assume in this section that the amplitudes / phases do not change with time.
	
	If $\nu_{\rm co}>\nu_{\rm res}$, then the wave's properties are unchanged within the bandwidth of the detector (such that $\langle E_{\rm i}E_{\rm i}^*\rangle_{\nu} \langle E_{\rm j}E_{\rm j}^*\rangle_{\nu}=\langle E_{\rm i}E_{\rm j}^*\rangle_{\nu} \langle E_{\rm j}E_{\rm i}^*\rangle_{\nu}$) and no depolarization will be measured.
	Consider instead that $\nu_{\rm co} < \nu_{\rm res}$ (ensuring that waves separated in frequency by at least $\nu_{\rm co}$ are integrated together by the detector). The result is $\langle E_{\rm i}E_{\rm i}^*\rangle_{\nu} \langle E_{\rm j}E_{\rm j}^*\rangle_{\nu}>\langle E_{\rm i}E_{\rm j}^*\rangle_{\nu} \langle E_{\rm j}E_{\rm i}^*\rangle_{\nu}$ leading to a depolarization of the signal. This case is the frequency domain equivalent of the $t_{\rm res}>t_{\rm var}$ case considered in \S \ref{sec:tempvar}. The degree of polarization as a function of the frequency ratio follows from the case of temporal variations: $\Pi_{\nu}=\Pi_t$.
	The coherence bandwidth can also be related to a coherence timescale $t_{\rm co}=\nu_{\rm co}^{-1}$. This is the time it takes waves separated by $\nu_{\rm co}$ to decohere. In cases where the spectral fluctuations are induced by a scintillating screen between the source and the observer, $t_{\rm co}$ can be related to the angular size of the observed screen, and the spatial scale of the inhomogeneities in it. Using this notation and recalling that by virtue of the uncertainty principle $\nu_{\rm res}\geq t_{\rm res}^{-1}$, we find $\nu_{\rm res}/\nu_{\rm co}\geq t_{\rm co}/t_{\rm res}$. Thus, $t_{\rm co}>t_{\rm res}$, is a sufficient (but not necessary) condition for depolarization due to frequency decoherence. 
	
	Combining the cases with both spectral and temporal variability, we conclude that depolarization is expected either for fine temporal resolution (and hence poor spectral resolution relative to the coherence bandwidth), $t_{\rm res}<t_{\rm co}$, or for poor temporal resolution relative to the intrinsic temporal variability $t_{\rm res}>t_{\rm var}$. 
	
	\subsection{Smoothly varying PA}
	\label{sec:meanRMPol}
	Another case of interest is one in which the PA of a wave is a continuous and ordered function of frequency.  A specific important example occurs when the wave passes through a column of magnetized plasma on the way to the observer, with some final rotation measure (RM). Faraday rotation then changes the wave's PA by a wavelength dependent amount: $\chi=\mbox{RM}\lambda^2$ (where $\lambda$ is the wave's wavelength). We define a characteristic frequency, $\nu_{\rm RM}$ such that the change in PA across $\nu_{\rm RM}$ is order unity
	\begin{equation}
		\label{eq:nuRM}
		\nu_{\rm RM}=\frac{\nu^3}{2\mbox{RM} c^2}
	\end{equation}
	where $\nu$ is the frequency at the center of the observed band and $c$ is the speed of light. As opposed to the stochastic case, the Stokes parameters no longer add up by random walk and in the limit $N_{\rm RM}=\nu_{\rm res}/\nu_{\rm RM}\gg1$ there is almost exact cancellation of oppositely directed PAs. The observed polarization is the small residual $\sim 1/N_{\rm RM}$ fraction of the signal that is not cancelled out. For general $N_{\rm res}$ we find
	\begin{equation}
		\label{eq:polRM}
		\Pi_{\rm RM}=\left(1+\frac{\nu_{\rm res}}{\nu_{\rm RM}}\right)^{-1}=\left[1+\left(\frac{\nu}{\nu_{\rm r,RM}}\right)^2\right]^{-1},
	\end{equation}
	where $\nu_{\rm r,RM}=\sqrt{2\mbox{RM} c^2\mathcal{R}}$ is defined such that $\nu_{\rm RM}(\nu_{\rm r,RM})=\nu_{\rm res}(\nu_{\rm r,RM})$ and in the second transition we have taken a concrete example in which the frequency resolution is proportional to the observed frequency and defined $\mathcal{R}\equiv \nu_{\rm res}/\nu$. 
	In the presence of fluctuations of the wave properties (either in the source or due to propagation), the analysis presented above can be applied to each coherent segment of length $\nu_{\rm co}$. The contributions of these independent segments is then combined randomly in a fashion similar to that described in \S \ref{sec:nuvar}.
	The resulting polarization is
	\begin{equation}
		\label{eq:combineRMstoch}
		\Pi_{\nu}=\left(1+\frac{\min(\nu_{\rm co},\nu_{\rm res})}{\nu_{\rm RM}}\right)^{-1} \left(1+\frac{\nu_{\rm res}}{\nu_{\rm co}}\right)^{-1/2}.
	\end{equation}
	Depending on the ordering of $\nu_{\rm co},\nu_{\rm res},\nu_{\rm RM}$, the effect of the stochasticity can be either to increase (when $\nu_{\rm RM}<\nu_{\rm co}<\nu_{\rm res}$) or decrease (when $\nu_{\rm co}\ll \min(\nu_{\rm res},\nu_{\rm RM})$) the polarization relative to the effect by the mean RM alone as quatified by equation \ref{eq:polRM}. 
	
	\section{A temporally variable wave passing through a scintillating screen}
	\label{sec:Screen}
	So far we have mainly focused on spectral or temporal variability in isolation. Next, we consider a model that combines the two effects. Our model consists of a temporally variable wave passing through a scintillating screen. The incident wave is decomposed as the sum of plane waves, and their complex amplitudes are assumed to be independent of frequency over the narrow frequency bandwidh of the detector.
	The discussion below summarized the important spatial and temporal scales in the system as well as the relationships between them (which are generally a function of frequency). Our discussion is roughly divided into a description of the important length scales and time scales in the system. A summary of the different temporal, spatial and spectral scales used in this work is given in table \ref{tbl:define} and a schematic sketch depicting the passage of a wave through a turbulent scattering screen is shown in figure \ref{fig:schem}.
	
	\subsection{Length scales associated with the screen}
	\label{sec:length}
	
	The dispersion relation for circularly polarized electromagnetic waves\footnote{A general elliptically polarized EM wave can be written as a linear combination of left- and right-circularly polarized waves with different phases.} in magnetized-plasma is:
	\begin{equation}
		c k_\pm = \omega - \frac{\omega_p^2}{2\omega}\left[ 1 \pm \frac{\omega_B}{\omega}\right],
		\label{EM-dispersion}
	\end{equation}
	where $\omega$ is the wave frequency, $k_\pm$ the wavenumbers for left \& right circularly polarized waves,
	\begin{equation}
		\omega_p = \sqrt{\frac{4\pi q^2 n_{\rm e}}{m}}, \quad\quad  \omega_B = \frac{q B_\parallel}{m c}
	\end{equation}
	are electron plasma and cyclotron frequencies, $q$ \& $m$ are electron charge and mass respectively, $n_{\rm e}$ is number density of electrons, and $B_{||}$ is the component of the magnetic field along the photon propagation direction.
	
	We consider how the polarization properties of a linearly polarized, coherent, EM wave are modified when it passes through a turbulent magnetized-plasma. The electric field vector at the observer after the wave has passed through the scattering screen can be calculated in the Fresnel diffraction limit, e.g. \cite{Narayan1992}
	\begin{equation}
		\label{eq:E_nu(x_0,t)}
		{\bf E_{\nu_0}(x_0, \rm t)} = {e^{i\omega t}\over R_{\rm F}^2} \int d^2x\, \exp\left\{{i\phi({\bf x}) + i\pi |{\bf x_0 - x}|^2/R_{\rm F}^2}\right\} {\bf E_{\nu_0}(x, \rm t_0)},
	\end{equation}
	where $t$ is the arrival time ($t_0$ is the arrival time from the center of the screen, ${\bf x}={\bf x_0}$), $R_{\rm F}=\sqrt{\lambda d}$ is the Fresnel radius, $d$ is the distance to the scattering screen from the source or the observer, whichever is smaller, and $\phi({\bf x})$ is the phase shift suffered by a linearly polarized wave as a result of non-zero $\omega_p$ as it crosses the screen at ${\bf x}$, which follows from the dispersion relation\footnote{The third term in the dispersion relation (eq. \ref{EM-dispersion}), $\pm\omega_p^2\omega_B/2\omega^2$, does not change the phase of a linearly polarized wave. It rotates the electric field vector of the linearly polarized wave by an amount $\chi$ given by eq. \ref{chi-mean}.} in eq. \ref{EM-dispersion}
	\begin{equation}
		\phi = \int ds \, {\omega_p^2\over 2c\omega}.
	\end{equation}
	And finally,
	\begin{equation}
		{\bf E_{\nu_0}(x, \rm t_0)} = {\bar{\bar T}}(\chi) {\bf E_{\nu_0,0}}
	\end{equation}
	is the electric field of the wave after exiting the screen at a point {\bf x} ({\bf E}$_{\nu_0,0}$ is the electric field just before the screen), ${\bar{\bar T}}(\chi)$ is a rotation matrix that rotates a vector by an angle $\chi$ given by
	
	\begin{equation}
		\chi = \int ds\, {\omega_p^2\omega_B\over 2c\omega^2} = {q^3\lambda^2\over 2\pi m^2 c^4}\int ds\, n_{\rm e}(s) B_\parallel(s) \equiv \lambda^2\,{\rm RM},
		\label{chi-mean}
	\end{equation}
	which can be easily derived from the dispersion relation (eq. \ref{EM-dispersion}). Both $n_{\rm e}$ and $B_{||}$ are likely to be inhomogeneous across the screen. The average phase change suffered by the wave after crossing the scattering screen due to its non-unity index of refraction is:
	\begin{equation}
		\phi_0 = \int ds\,\, {\omega_{\rm p}^2 \over2\omega c} = {q^2 ({\rm 1\, pc})\over \nu c m} [{\rm DM_{\rm s}}] = {2.6{\rm x}10^{7} [{\rm DM_{\rm s}}]\over \nu_9},
		\label{phi-mean}
	\end{equation}
	where ${\rm DM_{\rm s}}=n_{\rm e}L$/(1 pc) is the mean dispersion measure of the scattering screen; 1 pc $\equiv$ 3.1x10$^{18}$cm. A wave is scattered when passing through the plasma screen when the phase shift suffered by the wave fluctuates across the screen. 
	A particular case of interest for the density fluctuation is that of a Kolmogoroff spectrum of turbulent cascade. According to Kolmogoroff, the electron density fluctuation on scale $\ell$ is: $\delta n_{\rm e}(\ell)=n_{\rm e}(\ell/\ell_{\rm max})^{\alpha}$; where $\ell_{\rm max}$ is the maximum eddy size in the turbulent medium. In this case, the differential phase shift -- after subtracting the mean $\phi_0$ calculated above -- suffered by the wave after crossing the screen of thickness $L$, due to eddies of size $\ell$ is
	\begin{equation}
		\delta\phi = {q^2\over \nu c m}\int ds\, \delta n_{\rm e} \approx {\omega_{\rm p}^2\ell\over 2\omega c}\left[{L\over \ell}\right]^{1\over 2} \left[{\ell\over \ell_{\rm max}}\right]^{\alpha} = \phi_0 \left[{\ell\over L}\right]^{1\over 2} \left[{\ell\over \ell_{\rm max}}\right]^{\alpha}
		\label{delphi},
	\end{equation}
	where $\phi_0$ is given by equation (\ref{phi-mean}). Wave scattering in the screen is dominated by eddies of size $\ell_\phi$ which contribute to the phase fluctuation $\delta\phi\sim 1$. This important length scale can be calculated using equation \ref{delphi}  (for a more detailed description see e.g., \cite{BK2020}),
	\begin{equation}
		\label{eq:lpi}
		\ell_\phi \sim \phi_0^{-{2\over2\alpha+1}} \ell_{\rm max} \left({L\over \ell_{\rm max}}\right) ^{{1\over 2\alpha+1}} \propto \nu^{1.2}
	\end{equation}
	where the frequency dependence in the last part of the above expression is for $\alpha=1/3$ as appropriate for Kolmogoroff turbulence.
	
	In an analogous way, we calculate the different degree of rotation of the wave's electric field direction across the screen. The fluctuation to the rotation angle, $\delta\chi$, due to eddies of size $\ell$ stacked in a column in the screen along the line-of-sight to the observer is\footnote{We have assumed that $\delta n_{\rm e} \delta B_{||}$ is negligible compared to $B_{||}\delta n_{\rm e}$ and $n_{\rm e}\delta B_{||}$. This is equivalent to requiring that $\alpha>0$ \& $\beta>0$. The former is expected for a Kolmogoroff spectrum as mentioned above. The latter follows from the fact that large fluctuations in the magnetic field are more likely to be realized on greater scales, where magnetic tension and diffusion of the magnetic field lines (both of which work towards an ordering of the field lines) become weaker.}
	\begin{equation}
		\delta\chi = {q^3\lambda^2\over 2\pi m^2 c^4} \int ds\, \left[ n_{\rm e}\delta B_\parallel + B_\parallel \delta n_{\rm e}\right].
	\end{equation}
	We take fluctuations of $B_\parallel$ on scale $\ell$ to be given by the following powerlaw function \footnote{We assume here that the fluctuations in density and magnetic field are independent of each other. If the two were correlated this would lead to a reduction in the effective value of $\gamma$, and will generally cause the induced circular polarization due to multi path propagation to become more pronounced.}:
	\begin{equation}
		\delta B_{||}(\ell)=B_{||}(\ell/\ell_{\rm max})^{\beta}.
	\end{equation}
	Thus, ($n_{\rm e}\delta B_\parallel + B_\parallel \delta n_{\rm e}$) fluctuates on scale $\ell$ as $\propto (\ell/\ell_{\rm max})^{\gamma}$, where $\gamma \approx \min(\alpha,\beta)$. Making use of this scaling, the expression for $\delta\chi$ can be shown to reduce to
	\begin{equation}
		\delta\chi \approx \chi_0 \left[{\ell\over L}\right]^{1\over 2} \left[{\ell\over \ell_{\rm max}}\right]^{\gamma}
		\label{del-chi}
	\end{equation}
	where $\chi_0$, given by equation \ref{chi-mean}, is the mean rotation of the wave electric field direction by the magnetized screen. This can be expressed in terms of $\phi_0$ as follows
	\begin{equation}
		\chi_0 \approx \phi_0 \left\langle {\omega_B\over\omega}\right\rangle,
		\label{chi0}
	\end{equation}
	with $\langle\omega_B\rangle\equiv q \langle B_\parallel\rangle/(m c)$ being the mean cyclotron frequency in the scattering screen. 
	
	We define $\ell_{\chi}$ as the size of eddies for which $\delta\chi \sim 1$. The expression for $\ell_{\chi}$ follows from equation (\ref{del-chi})
	\begin{equation}
		\label{eq:lchi}
		\ell_{\chi} \sim \chi_0^{-{2\over 2\gamma+1}} \ell_{\rm max} \left(\frac{L}{\ell_{\rm max}}\right)^{1\over 2\gamma+1}\!\propto \! \nu^{2.4}, \quad\quad \gamma=\min(\alpha,\beta), 
	\end{equation}
	where the dependence of $\ell_\chi$ on frequency is for $\alpha=\beta=1/3$. The ratio $\ell_\chi/\ell_\phi$ can be obtained by combining equations (\ref{eq:lpi}), (\ref{chi0}) \& (\ref{eq:lchi})
	\begin{equation}
		{\ell_\chi\over\ell_\phi} \sim \left[{\omega\over\omega_B}\right]^{2\over 2\gamma+1} \left( {L\over\phi_0^2\,\ell_{\rm max} }\right)^{2(\alpha-\gamma)\over (2\alpha+1)(2\gamma+1)}.
	\end{equation}
	This expression simplifies considerably for the special case of $\beta=\alpha$ (leading to $\gamma=\alpha$):
	
	\begin{equation}
		{\ell_\chi\over\ell_\phi} \sim \left[{\omega\over\omega_B}\right]^{2\over 2\alpha+1} \sim \left( {2.6{\rm x}10^7  [{\rm DM_s}]\over \nu_9\,\chi_0}\right)^{2\over 2\alpha+1}.
		\label{lchi_lphi}
	\end{equation}
	
\begin{figure*}
		\centering
		\includegraphics[width = 0.4\textwidth]{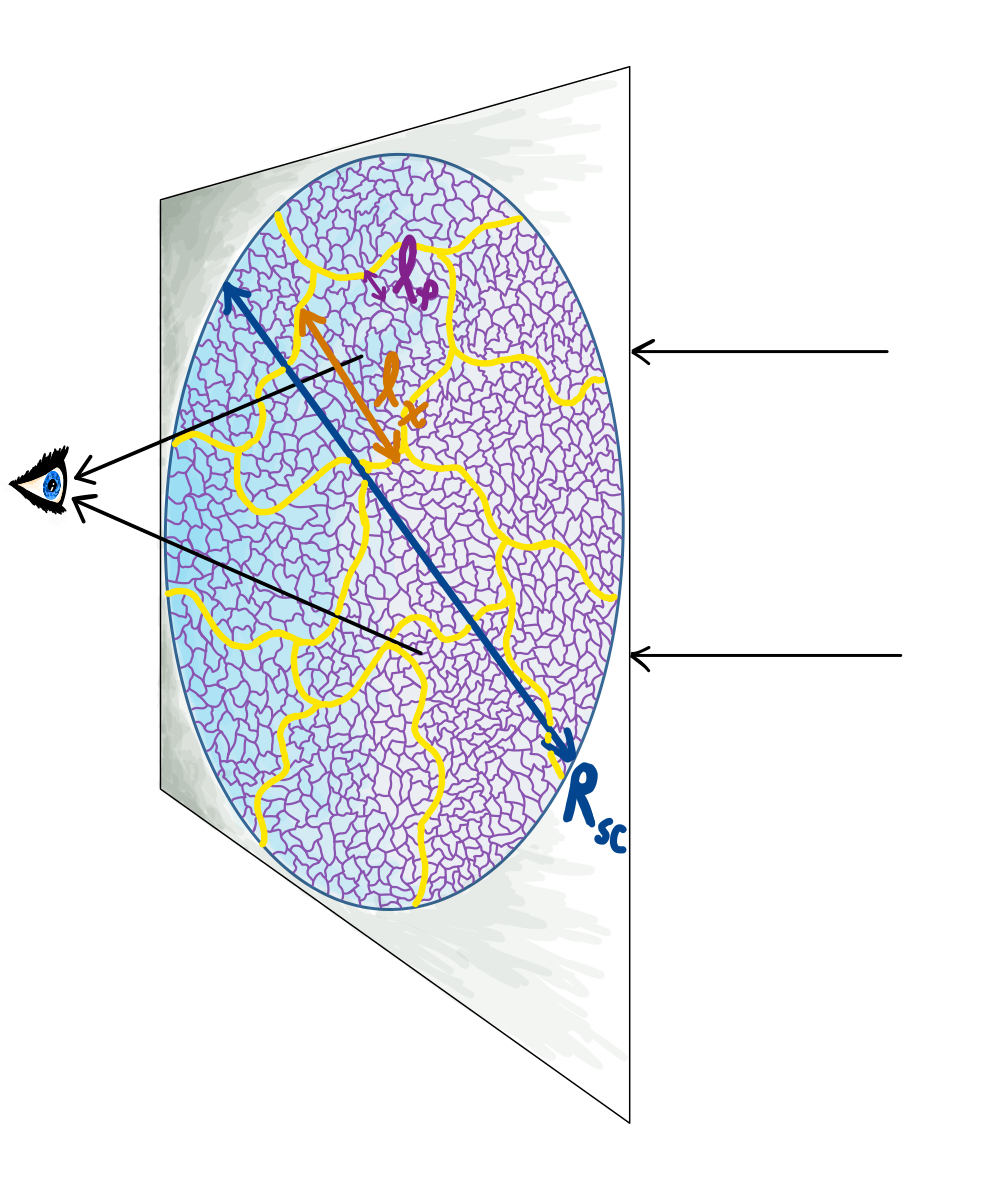}
		\caption{Schematic sketch depicting the passage of a wave through a turbulent scattering screen, which changes the phase of the incoming wave (at its central frequency) on a spatial scale $\ell_{\phi}$ and its PA on a spatial scale $\ell_{\chi}$. For most cases of interest $\ell_{\chi}\gg \ell_{\phi}$.}
		\label{fig:schem}
	\end{figure*}

	For a scattering screen with mean magnetic field of strength 10 $\mu$G, and observing frequency of 1 GHz, $\omega/\omega_B=10^8$. Therefore, $\ell_\chi/\ell_\phi \sim 10^{10}$ for $\alpha=1/3$. Thus, for most physical parameters of interest for ISM scattering screens $\ell_\chi \gg\ell_\phi$. However, it should be noted that for FRB 20121102A, RM $\sim 10^5\mbox{rad m}^{-2}$, and thus $\chi_0\sim 10^4$ at 1 GHz. If the plasma screen responsible for this large rotation measure has DM$_{\rm s}\sim 1$ pc cm$^{-3}$, then in that case $\ell_\chi/\ell_\phi\sim 10^4$, and very likely $\ell_\chi$ is smaller than the segment of the screen from which waves are scattered toward the observer. The implications of this for FRB 20121102A, given its high observed degree of linear polarization, are discussed in \S \ref{obs}.
	A sketch representing the scales $\ell_{\phi},\ell_{\chi}$ in the scattering screen are shown in figure \ref{fig:schem}.
	
	The requirement that $\ell_{\chi}<\ell_{\rm max}$, a necessary condition for the screen to induce stochastic depolarization, leads to (see eq. \ref{eq:lchi})
	\begin{equation}
		\label{eq:chi0}
		\chi_0>\left(\frac{L}{\ell_{\rm max}}\right)^{1/2}\geq 1
	\end{equation}
	and that means that the mean PA rotation induced by the screen must be large for stochastic Faraday depolarization. As we describe below, this is a necessary but not sufficient condition for the screen to cause depolarization and / or induce circular polarization, which requires $\ell_{\chi}\lesssim R_{\rm sc}$ (where $R_{\rm sc}$, the scattering radius defined in equation \ref{eq:Rsc} below, is the size of the scattering screen from which photons are scattered in the direction to the observer). The advantage of using the weaker condition imposed by equation \ref{eq:chi0}, is the simplicity of its application to data as it relates directly to a single observed parameter. 
	
	Nevertheless, it should be emphasized that although $\ell_{\chi}$ is associated with fluctuations in $n_{\rm e}$ and $B_{||}$ (which are smaller than the mean values of those parameters), the stochastic rotation of wave electric vector across the screen is not a second order effect as compared to the mean rotation introduced by the screen. This is demonstrated in equation \ref{eq:combineRMstoch} which shows that when stochastic effects are taken into account, the polarization degree can change significantly relative to the polarization of a wave with fixed properties passing through a medium with uniform RM (see also figure \ref{fig:polarizationvsmean}). Furthermore, as we will show in detail in \S \ref{sec:toymodelPi}, \ref{sec:Polscreen}, the stochastic nature of the screen also induces a large circularly polarized component, even if the incoming wave is completely linearly polarized. 
	Furthermore, as we outline in \S \ref{sec:discuss}, there are several techniques that could be used to remove the depolarization caused by the mean properties of the screen.
	
	The effective eddy size that determines the wave deflection is set by $\ell_\phi$ as there is no phase change associated with the Faraday rotation of the wave electric field vector. Waves are deflected by the $\ell_{\phi}$ patches of the screen by an angle $\delta\theta \sim \lambda/\pi\ell_{\phi}$.
	
	The radius of the screen, $R_{\rm screen}$, from which the observer receives the radiation is the minimum of $L$ (the physical size of the turbulent screen) and the scattering radius
	\begin{equation}
		\label{eq:Rsc}
		R_{\rm sc} \sim (d\delta\theta)\sim \frac{R_{\rm F}^2}{\pi\ell_{\phi}}
	\end{equation}
	For most of the astrophysically relevant parameter space, we find that $R_{\rm sc}\ll L$. We therefore focus on this case in what follows. The condition for strong scattering regime, $R_{\rm F}\gg \ell_{\phi}$ or $R_{\rm sc}\gg R_{\rm F}$, is readily realized at lower frequencies since $R_{\rm F}/\ell_{\phi}\propto \nu^{-1.7}$.

	As we discuss in \S \ref{sec:Polscreen} in order for the screen to induce a circularly polarized component in the observed wave (and to potentially lead to stochastic depolarisation), a required condition is $\ell_{\chi}<R_{\rm sc}$.
	We therefore define an additional characteristic frequency, $\nu_{\chi \rm s}$ such that $\ell_{\chi}(\nu_{\chi \rm s})=R_{\rm sc}(\nu_{\chi \rm s})$,
	\begin{eqnarray}
		\label{eq:nuchis}
		&\nu_{\chi\rm s}=\left(\frac{cd}{\pi}\right)^{(2\gamma+1)(2\alpha+1)\over 4\alpha\gamma+10\alpha+6\gamma+7}\left( {m c  \over q^2 n_{\rm e} L^{1/2}}\right)^{-4(\alpha+\gamma+1)\over 4\alpha\gamma+10\alpha+6\gamma+7} \\ & \times  \left( {2\pi m c \over qB_{||}}\right)^{-2(2\alpha+1)\over4\alpha\gamma+10\alpha+6\gamma+7} \ell_{\rm max}^{-{2(4\alpha\gamma+\alpha+\gamma)\over 4\alpha\gamma+10\alpha+6\gamma+7}}\nonumber
	\end{eqnarray}
	It is useful to define a frequency, $\nu_{\rm *}$ for which $\ell_{\phi}(\nu_*)=R_{\rm sc}(\nu_*)$
	\begin{eqnarray}
		\label{eq:nustar}
		\nu_*=\left(\frac{cd}{\pi}\right)^{2\alpha+1\over 2\alpha+5} \left( {m c \over q^2 n_{\rm e} }\right)^{-4\over 2\alpha+5}\ell_{\rm max}^{-4\alpha\over 2\alpha+5}L^{2\over 2\alpha+5}
	\end{eqnarray}
	and which determines the scattering regime.

\begin{figure*}
		\centering
		\includegraphics[width = 0.4\textwidth]{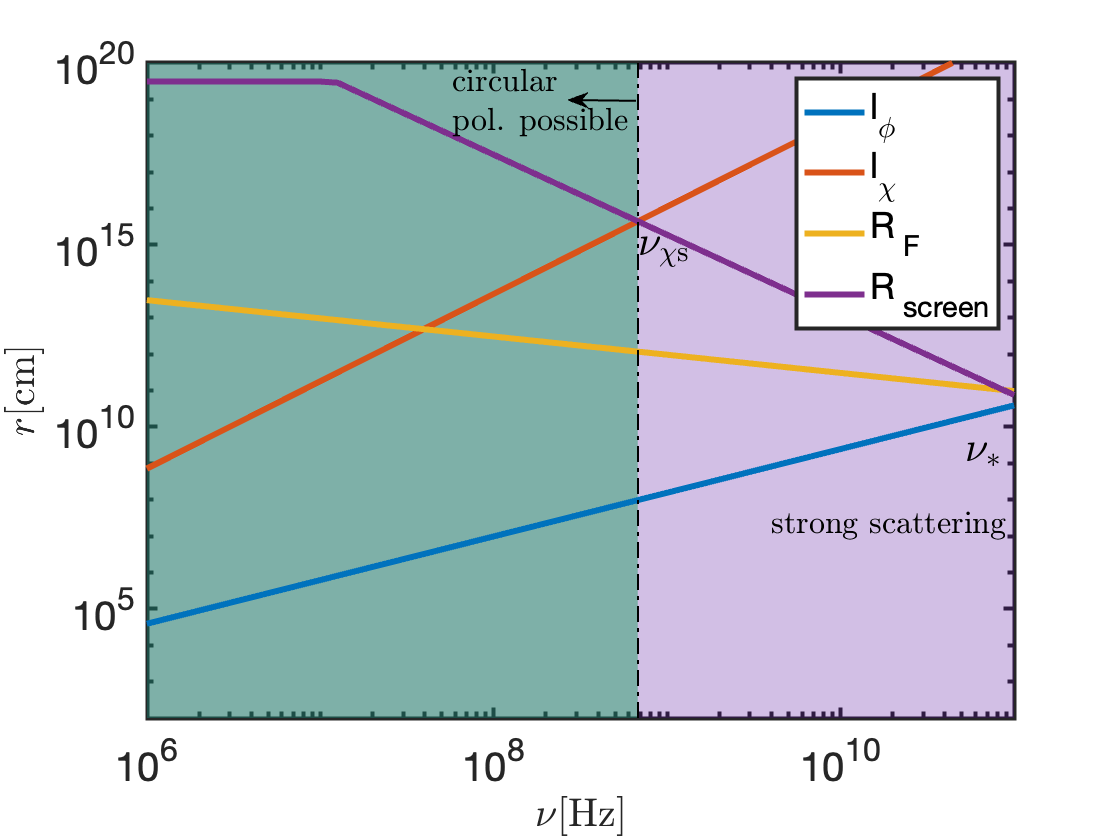}
		\includegraphics[width = 0.4\textwidth]{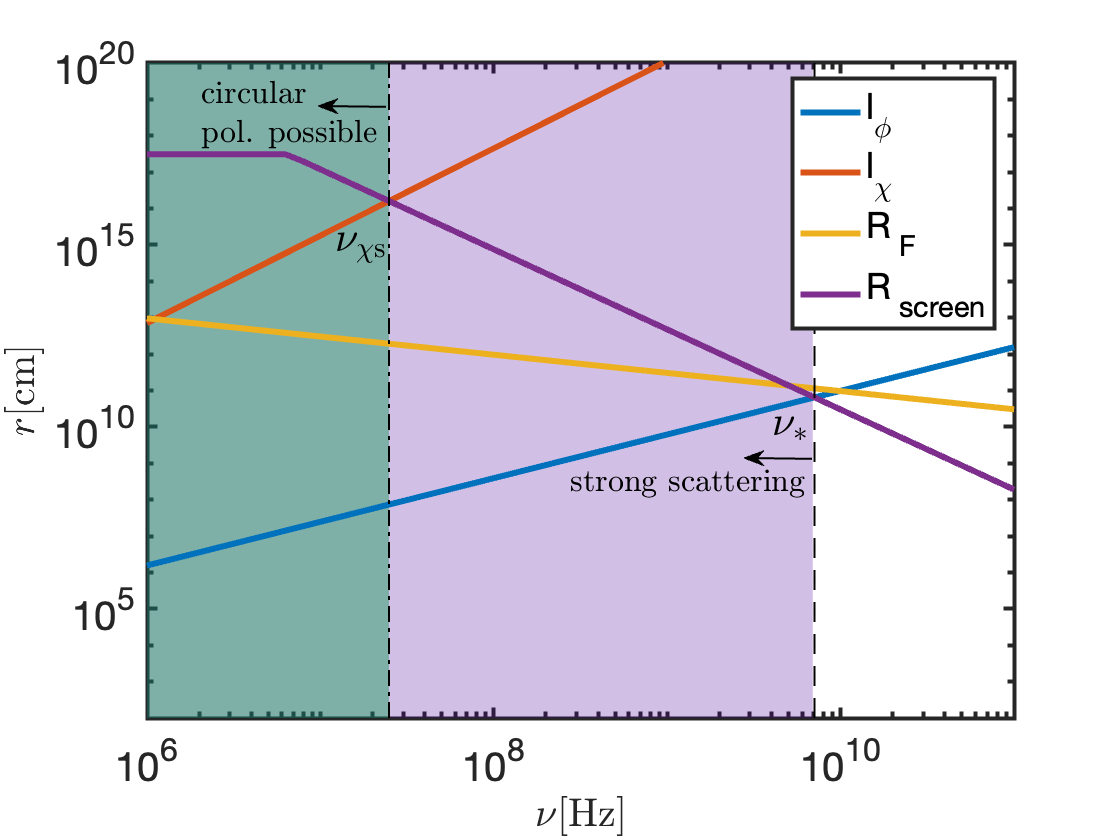}
		\caption{Scattering screen spatial scales as a function of frequency for a screen in an ISM-like environment. Parameters assumed for this calculation are, left panel: $\ell_{\rm max}=L=10\mbox{pc},d=10\mbox{kpc},B_{||}=100\mu\mbox{G},n=10\mbox{cm}^{-3}, \alpha=\beta=1/3$ and right panel: $\ell_{\rm max}=L=0.1\mbox{pc},d=1\mbox{kpc},B_{||}=\mu\mbox{G},n=1\mbox{cm}^{-3}, \alpha=\beta=1/3$. These parameters lead to RM$_{\rm s}=8\times 10^3\mbox{rad m}^{-2}$ (left panel) and RM$_{\rm s}=8\mbox{rad m}^{-2}$ for the right panel.}
		\label{fig:spatialscales}
	\end{figure*}

	The frequency dependence of the different spatial scales is summarized in table \ref{tbl:spatial} and estimated for a couple of MW-type scintillating screens in figure \ref{fig:spatialscales}  (note that contributions from scattering in the IGM are likely to be sub-dominant, see \S 3 of \cite{BK2020} for details). For typical FRB observations at $\sim 100\mbox{MHz}-10\mbox{GHz}$ we find that (especially towards the lower end of the observed frequency range) the strong scattering limit is often applicable. Furthermore, we find that $\ell_{\chi}$ is likely to be orders of magnitude greater than $\ell_{\phi}$ (see eq. \ref{lchi_lphi}). At the same time, at the lower end of the observed frequency range, it is reasonable to expect $\ell_{\chi}<R_{\rm sc}$, leading to significant circular polarization induced by the screen.

	\begin{table}
		\centering
		\caption{\label{tbl:define}
			Temporal, spatial and frequency scales used in this work.}
		\centering
		\resizebox{0.48\textwidth}{!}{
			\begin{threeparttable}	\begin{tabular}{ccc}\hline	
					Notation & Definition & Equation\tabularnewline
					\hline 
					$T$  & wave period  & -\tabularnewline
					$t_{\rm res}$  & detector temporal resolution  & -\tabularnewline
					$t_{\rm res,lim}$  & limiting temporal resolution & $\nu_{\rm res}^{-1}$ \tabularnewline
					$t_{\rm scat}$  & delay associated with scattering screen  & \ref{eq:tscat}\tabularnewline
					$t_{\rm var}$  & source temporal variability  & -\tabularnewline
					$t_{\rm scr,var}$  & temporal variability due to screen  & \ref{eq:tscrvar}\tabularnewline
					$t_{\rm var,eff}$  &  effective variability time determining $\Pi_{\rm lin}$ & $\min(t_{\rm var},t_{\rm scr,var})$ \tabularnewline
					\hline
					$\ell_{\rm max}$  & maximum size of eddies in screen & - \tabularnewline
					$L$  & size of turbulent screen & - \tabularnewline
					$\ell_{\phi}$  & size of eddies dominating phase shift  & \ref{eq:lpi}\tabularnewline
					$\ell_{\chi}$  & size of eddies dominating field rotation  & \ref{eq:lchi} \tabularnewline
					$R_{\rm F}$  & Fresnel radius & $\sqrt{\lambda d_{\rm l}}$ \tabularnewline
					$R_{\rm sc}$  & scattering radius & \ref{eq:Rsc} \tabularnewline
					$R_{\rm screen}$  & visible radius of screen & $\min(L,R_{\rm sc})$ \tabularnewline
					\hline
					$\nu_0$  & central observed frequency & - \tabularnewline
					$\nu_{\rm res}$  & spectral resolution & - \tabularnewline
					$\nu_{\rm RM}$  & band of PA change due to non-turbulent screen & \ref{eq:nuRM} \tabularnewline
					$\nu_{\rm r,RM}$  & $\nu_{\rm RM}(\nu_{\rm r,RM})=\nu_{\rm res}(\nu_{\rm r,RM})$ &$\sqrt{2\mbox{RM}c^2\mathcal{R}}$ \tabularnewline
					$\nu_{\rm co}$  & coherence bandwidth & $t_{\rm scat}^{-1}$ \tabularnewline
					$\nu_{\chi \rm s}$  & $\ell_{\chi}(\nu_{\chi \rm s})=R_{\rm sc}(\nu_{\chi \rm s})$ & \ref{eq:nuchis} \tabularnewline
					$\nu_*$  & $\ell_{\phi}(\nu_*)=R_{\rm sc}(\nu_*)$ & \ref{eq:nustar} \tabularnewline
					$\nu_{\rm var}$ & $t_{\rm var}=t_{\rm scr,var}(\nu_{\rm var})$ & \ref{eq:nuvar} \tabularnewline
					$\nu_{\rm rs}$ & $t_{\rm scat}(\nu_{\rm rs})=t_{\rm res,lim}(\nu_{\rm rs})$ & \ref{eq:nurs} \tabularnewline
					$\nu_{\rm rv}$ & $t_{\rm var,eff}(\nu_{\rm rv})=t_{\rm res,lim}(\nu_{\rm rv})$ & \ref{eq:nurv} \tabularnewline
					$\nu_{\rm N}$ & $N_{\chi}(\nu_{\rm N})=N_{\rm co}(\nu_{\rm N})$ & \ref{eq:nuN} \tabularnewline
				\end{tabular}
			\end{threeparttable}
		}
	\end{table} 
	
	\subsection{Time scales associated with the screen}
	\label{sec:timescales}
	In the strong scattering limit, the effective eddy size, $\ell_{\phi}$, is very small compared to the size of the screen the observer sees, $R_{\rm screen}$. This suggests that scatterings can lead to significant temporal variability (see \citealt{BK2020} for details). The variability time associated with the eddy turnover time or the screen's motion relative to the observer is given by
	\begin{equation}
		\label{eq:tscrvar}
		t_{\rm scr,var}\approx \ell_{\phi}/\sqrt{v_{\rm ed}^2+v_{\rm os}^2}\propto \nu^{2\over 2\alpha+1}
	\end{equation}
	where $v_{\rm ed}$ is the eddy speed and $v_{\rm os}$ is the transverse velocity between the scattering screen and the observer.
	The intrinsic variability of the source, $t_{\rm var}$, is unlikely to be strongly frequency dependent. The effective variability time for the purpose of determining the {\it linear} polarization only (see \S \ref{sec:toymodelPi},\ref{sec:Polscreen}) is the shorter of these two timescales $t_{\rm var,eff}=\min(t_{\rm var},t_{\rm scr,var})$. We define a characteristic frequency $\nu_{\rm var}$ such that $t_{\rm var}=t_{\rm scr,var}(\nu_{\rm var})$
	\begin{eqnarray}
		\label{eq:nuvar}
		\nu_{\rm var}=(v_{\rm max}t_{\rm var})^{2\alpha+1\over 2} \ell_{\rm max}^{-\alpha}L^{1/2} \left({q^2 n_{\rm e} \over m c}\right)
	\end{eqnarray}
	where $v_{\rm max}=\max(v_{\rm ed},v_{\rm os})$.
	
	The scattering screen averages the EM pulse on a timescale
	\begin{eqnarray}
		\label{eq:tscat}
		t_{\rm scat}=\frac{R_{\rm sc}^2}{2d c}=\frac{T}{2\pi^2}\frac{R_{\rm F}^2}{\ell_{\phi}^2}\propto \nu^{-(4\alpha+6)\over 2\alpha+1} 
	\end{eqnarray}
	In particular, for $\alpha=1/3$, $t_{\rm scat}\propto \nu^{-4.4}$.
	
\begin{figure*}
		\centering
		\includegraphics[width = 0.4\textwidth]{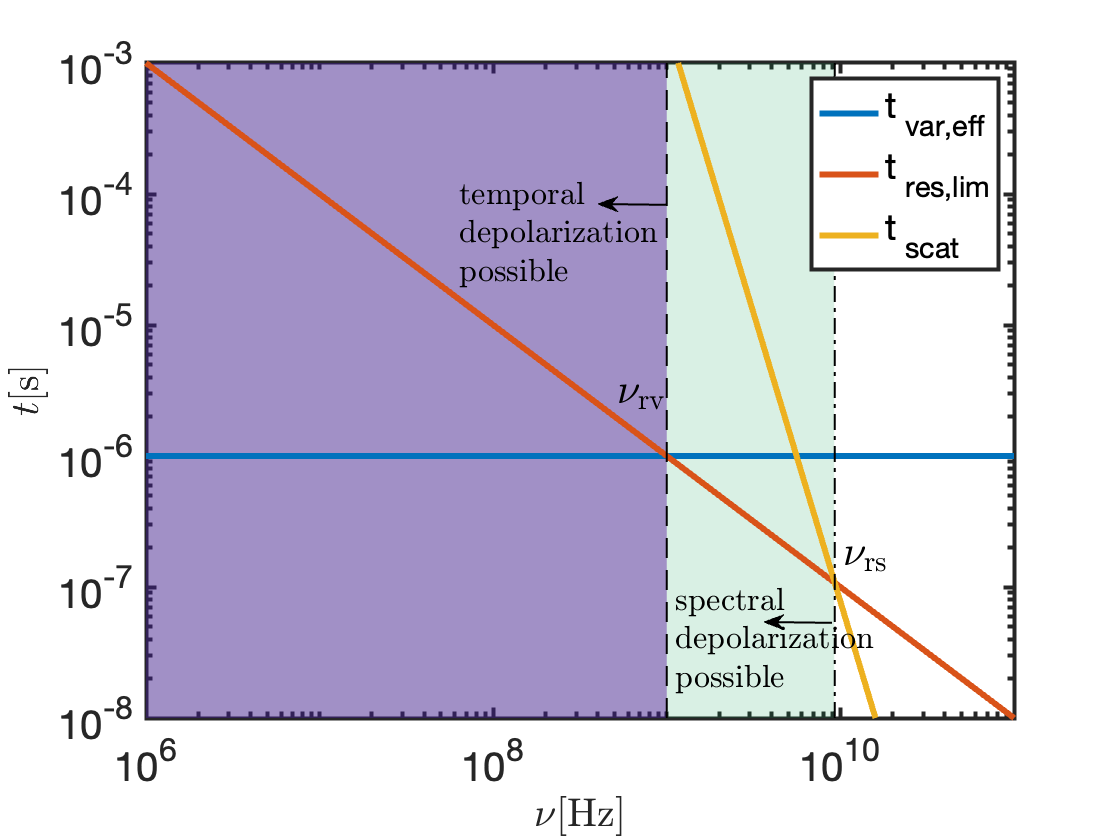}
		\includegraphics[width = 0.4\textwidth]{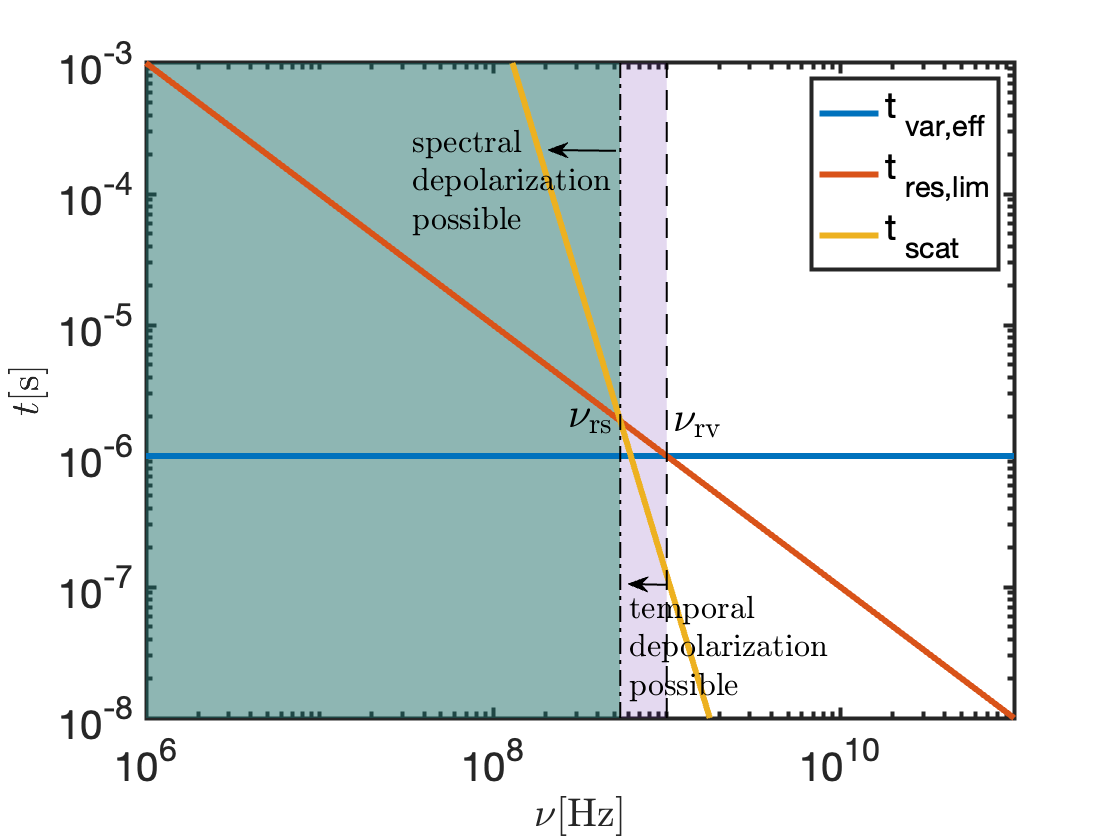}
		\caption{Temporal scales as a function of frequency. Parameters assumed for this calculation are the same as in figure \ref{fig:spatialscales} with the addition of: $v_{\rm max}=100\mbox{km s}^{-1}, t_{\rm var}=1\mu\mbox{s},\mathcal{R}=10^{-3}$. In both cases, $t_{\rm scr,var}\gg t_{\rm res,lim},t_{\rm var}$ for the entire range of frequencies depicted. In this situation "temporal depolarization" holds only for the linear component of the polarization, as described in \S \ref{sec:toymodelPi}.}
		\label{fig:tempscales}
	\end{figure*}

	Finally, due to the uncertainty principle, the limiting temporal resolution of a detector scales as $t_{\rm res,lim}\propto \nu^{-1}$ (the actual temporal resolution, $t_{\rm res}$ satisfies $t_{\rm res}\geq t_{\rm res,lim}$). Similar to the discussion regarding the spatial scales, we can define two additional characteristic frequencies. The first, $\nu_{\rm rs}$, is defined such that $t_{\rm scat}(\nu_{\rm rs})=t_{\rm res,lim}(\nu_{\rm rs})$
	\begin{eqnarray}
		\label{eq:nurs}
		\nu_{\rm rs}=\left({c d \mathcal{R} \over 2\pi^2}\right)^{2\alpha+1\over 2\alpha+5} \ell_{\rm max}^{-4\alpha\over 2\alpha+5}L^{2\over 2\alpha+5} \left({m c \over q^2 n_{\rm e}}\right)^{-4\over 2\alpha+5}
	\end{eqnarray}
	where $\mathcal{R}\equiv \nu_{\rm res}/\nu$ (such that $t_{\rm res,lim}=(\nu\mathcal{R})^{-1}$).
	A second frequency, $\nu_{\rm rv}$, is defined such that $t_{\rm var,eff}(\nu_{\rm rv})=t_{\rm res,lim}(\nu_{\rm rv})$
	\begin{eqnarray}
		\label{eq:nurv}
		\nu_{\rm rv}=\max\bigg[(\mathcal{R}t_{\rm var})^{-1},\left({v_{\rm max}\over \mathcal{R}}\right)^{2\alpha+1\over 2\alpha+3} \left({m c \over q^2 n_{\rm e}}\right)^{-2\over 2\alpha+3}\ell_{\rm max}^{-2\alpha\over 2\alpha+3}L^{1\over 2\alpha+3}\bigg]
	\end{eqnarray}
	We note that when $\nu\ll \nu_{\rm rs}$, we have $t_{\rm scat}\gg t_{\rm res,lim}$. This is equivalent to $\nu_{\rm co}\ll \nu_{\rm res}$ and means that the wave is depolarized due to finite spectral resolution (see \S \ref{sec:nuvar}). At high frequencies -- $\nu\gg \nu_{\rm rs}$ $\rightarrow$ $\nu_{\rm co}\gg \nu_{\rm res}$ -- no spectral depolarization occurs. Similarly at $\nu\ll \nu_{\rm rv}$ we expect $t_{\rm res,lim}\gg t_{\rm var,eff}$. This limit corresponds to depolarization due to the finite temporal resolution of the detector (see \S \ref{sec:tempvar}). In the opposite limit, if $t_{\rm res}=t_{\rm res,lim}$, then $\nu\gg\nu_{\rm rv}$, which corresponds to $t_{\rm res}\ll t_{\rm var,eff}$, and no temporal depolarization occurs. The results are summarized in table \ref{tbl:spatial} and a specific example for a possible frequency evolution is shown in figure \ref{fig:tempscales}. We conclude that at sufficiently low frequencies, we have $t_{\rm var,eff}<t_{\rm res}<t_{\rm scat}$ while at sufficiently high frequencies $t_{\rm scat}<t_{\rm res}<t_{\rm var,eff}$. As we expand on in \S \ref{sec:Polscreen} the implication is that for lower observed frequencies the signal is strongly depolarized (due to both frequency and temporal decoherence) while at higher frequencies the wave maintains its original polarization.
	
\begin{figure*}
		\centering
		\includegraphics[width = 0.4\textwidth]{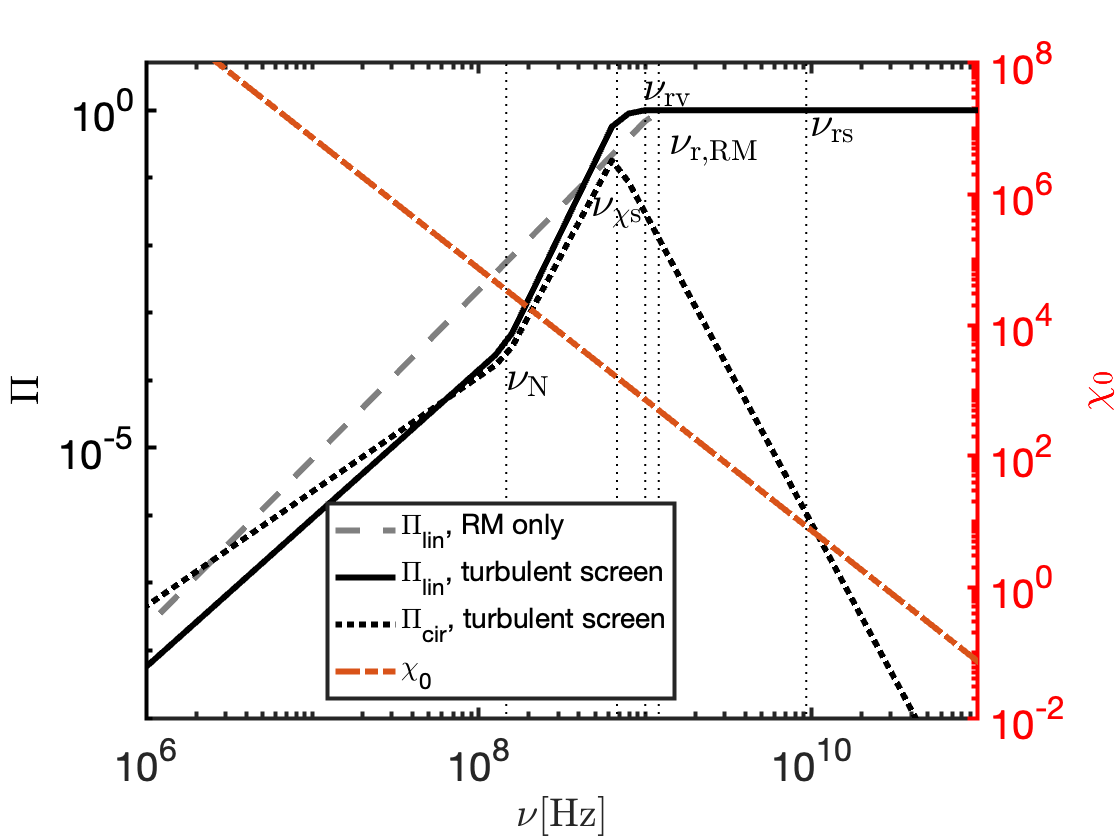}
		\includegraphics[width = 0.4\textwidth]{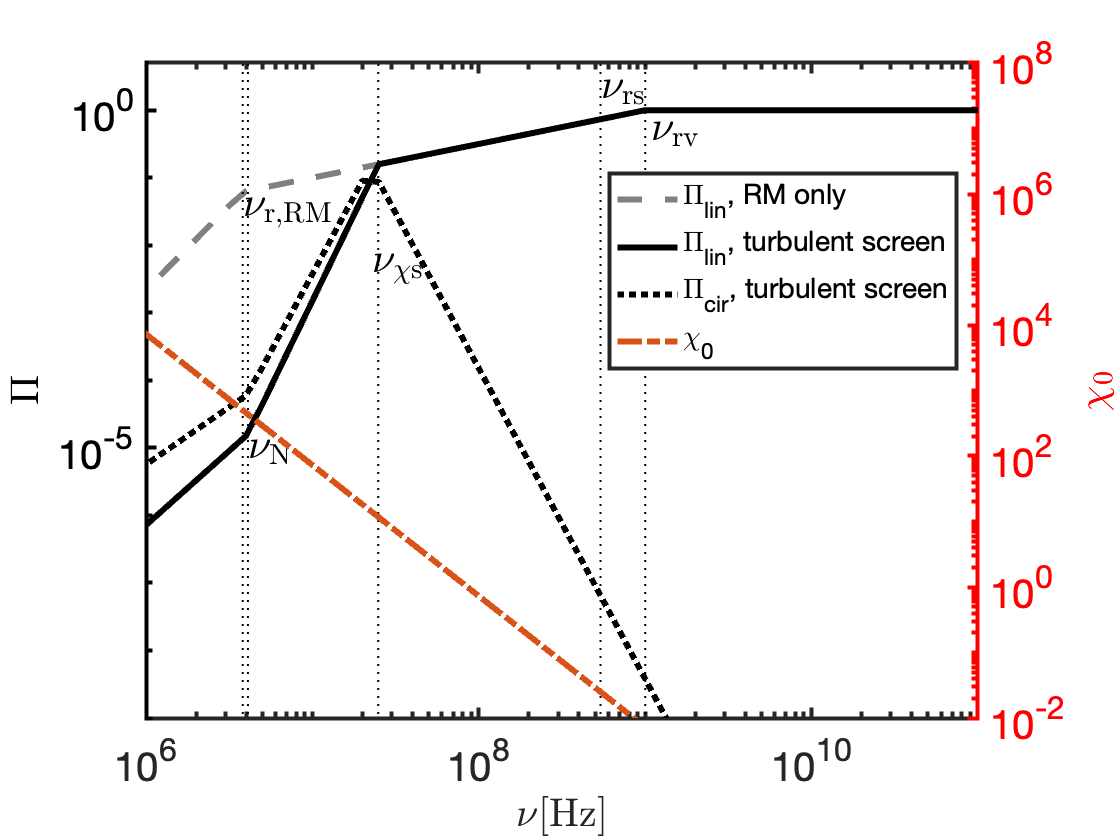}
		\caption{ A solid (dotted) line depicts the resulting linear (circular) polarization for an initially linearly polarized wave passing through screens with the same properties as described in figures \ref{fig:spatialscales}, \ref{fig:tempscales} (given by equation \ref{eq:Pisummarizetot}). As a comparison we plot also with a dashed line, the linear polarization resulting from passage of the same wave through single column (as opposed to multi-path propagation, here the circular polarization remains zero at the detector) with a rotation measure equal to the mean value produced by the screen (given by equation \ref{eq:combineRMstoch}). We assume the source to have the same temporal variability in both cases, and to have no significant spectral variability across the observed bandwidth.}
		\label{fig:polarizationvsmean}
	\end{figure*}

	Finally, for the same screen, source and detector properties plotted in figures \ref{fig:spatialscales}, \ref{fig:tempscales} we plot in figure \ref{fig:polarizationvsmean} the level of observed polarization as compared to the polarization due to a non stochastic column with the same mean RM. As shown in \S \ref{sec:meanRMPol}, in order for the mean RM of the screen to affect the polarization, we must have $\min(\nu_{\rm co},\nu_{\rm res})>\nu_{\rm RM}$ (where for a scintillating screen, $\nu_{\rm co}=t_{\rm scat}^{-1}$, see \S \ref{sec:Polscreen} for details). Typically, we have $\nu_{\rm res}\propto \nu, \nu_{\rm RM}\propto \nu^3, \nu_{\rm co}\propto \nu^{4.4}$. Since the frequency dependence of $\nu_{\rm RM}$ is between the other two frequencies, the result is that if $\nu_{\rm r,RM}<\nu_{\rm rs}$ then $\min(\nu_{\rm co},\nu_{\rm res})<\nu_{\rm RM}$ for all frequencies and the mean RM of the screen never affects the observed polarization. This ordering of the frequencies is very natural for ISM-like screens. Indeed, as shown in figure \ref{fig:polarizationvsmean}, $\nu_{\rm r,RM}\ll \nu_{\rm rs}$ for both the examples considered. Therefore the depolarization is entirely dominated by the stochastic nature of the screen and is unaffected by the screen's mean RM. This motivates us to investigate in detail the level of depolarization caused by the stochastic screen, which is explored next.

	\begin{table}
		\centering
		\caption{\label{tbl:spatial}
			Frequency regimes for scattering spatial and temporal scales. We have assumed here $\alpha=1/3,\beta=0$ and $t_{\rm res}=t_{\rm res,lim}$. The classification remains the same for other values of those parameters but the appropriate regime for $\nu_0<\nu_X$ and $\nu_0>\nu_X$ may be flipped if the values of these parameters significantly deviate from those written above.}
		\centering
		\resizebox{0.45\textwidth}{!}{
			\begin{threeparttable}	\begin{tabular}{ccc}\hline	
					$\nu_X$ & $\nu_0<\nu_X$ & $\nu_0>\nu_X$\tabularnewline
					\hline 
					$\nu_{*}$  & strong scattering  & weak scattering\tabularnewline
					$\nu_{\chi \rm s}$  & circular pol. possible & no circular pol.\tabularnewline
					$\nu_{\rm rs}$  & spectral depolarization & no spectral depolarization \tabularnewline
					$\nu_{\rm rv}$  & temporal depolarization & no temporal depolarization\tabularnewline
					\hline
				\end{tabular}
			\end{threeparttable}
		}
	\end{table} 
	
	\subsection{Toy model for polarization of a wave passing through scintillating screen with arbitrary RM}
	\label{sec:toymodelPi}
\begin{figure}
		\centering
		\includegraphics[width = 0.4\textwidth]{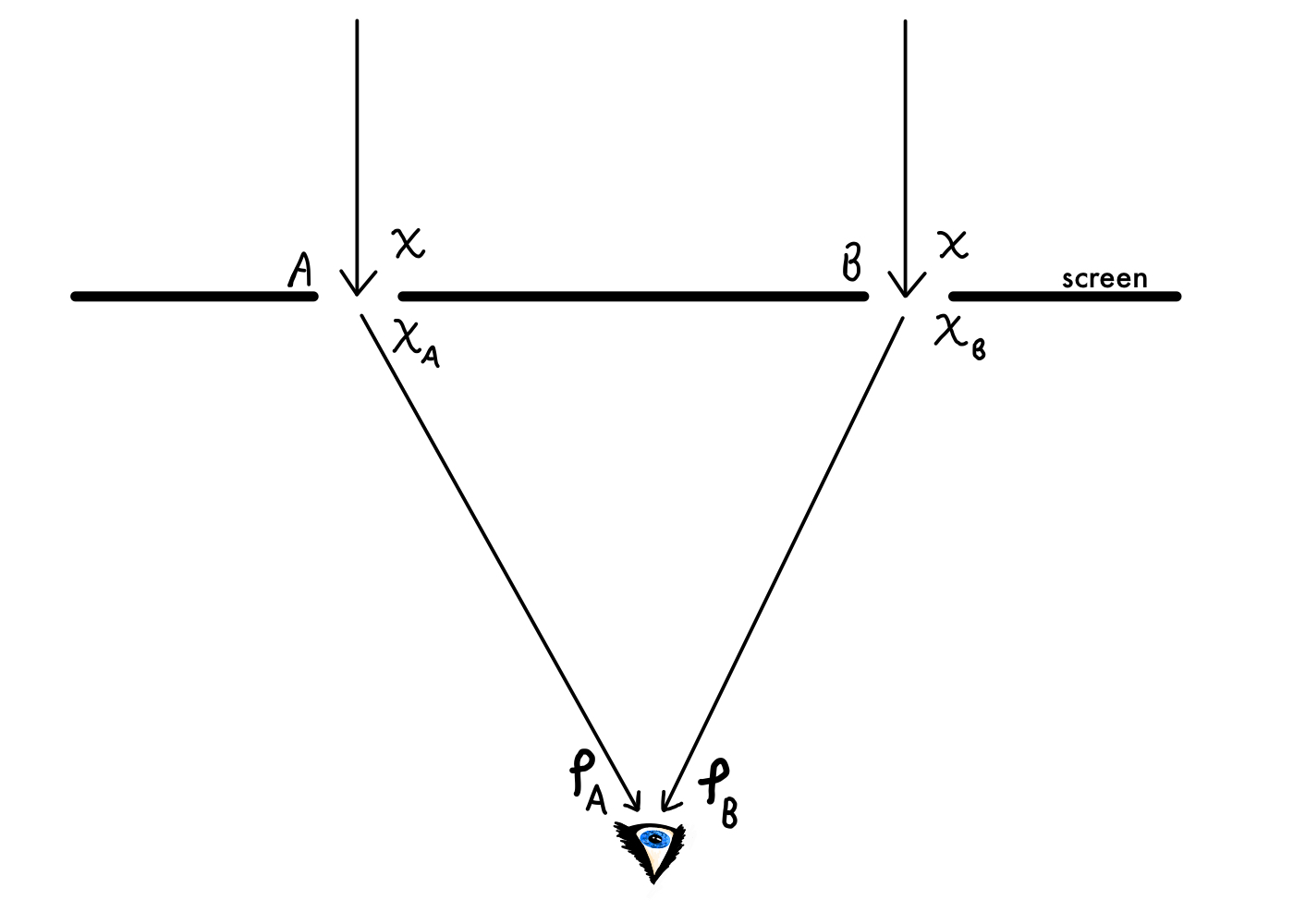}
		\caption{A linearly polarized wave (with PA=$\chi$) going through a screen with two slits ($A,B$). The PA of the wave components outgoing from the two slits are $\chi_{\rm A},\chi_{\rm B}$. The accumulated phases of the wave at the observer after crossing the slits $A$ and $B$ are $\phi_{\rm A},\phi_{\rm B}$; the phase change includes the contribution due to non-unity index of refraction of the plasma in the screen, and the difference in optical path length.}
		\label{fig:twoslit}
	\end{figure}

	Before turning to the polarization signature of a wave passing through a turbulent screen, it is useful to consider an instructive toy model that allows us to explicitly demonstrate the decrease in observed spectral polarization for a wave passing through a turbulent screen (the effects of temporal depolarization are simpler as described in \S \ref{sec:tempvar}, and can be decoupled from the spectral depolarization, shown in figure \ref{fig:1D} below). The key aspect introduced by scintillation is that the EM fields recorded by an observer are a superposition of scattered waves from different patches on the screen which induce different phases and PA changes to the incoming wave. As it turns out, a superposition of two such components (denoted here with subscripts A and B, see figure \ref{fig:twoslit}) can capture the key features of a magnetized scintillation screen, and we therefore focus on this situation in what follows. 
	For clarity, we begin by considering an idealized case in which the detector has an infinitely narrow spectral and temporal resolution. The dimensionless electric field can then be written as
	\begin{eqnarray}
		&   E_1(\tilde{\nu})=\frac{1}{\sqrt{2}}\left[\cos(\chi_{\rm A}/\tilde{\nu}^2)e^{i\phi_{\rm A}}+\cos(\chi_{\rm B}/\tilde{\nu}^2)e^{i\phi_{\rm B}} \right]\\
		&   E_2(\tilde{\nu})=\frac{1}{\sqrt{2}}\left[\sin(\chi_{\rm A}/\tilde{\nu}^2)e^{i\phi_{\rm A}}+\sin(\chi_{\rm B}/\tilde{\nu}^2)e^{i\phi_{\rm B}}\right]
	\end{eqnarray}
	where $\vec{E}=\hat{x}E_1+\hat{y}E_2$, $\tilde{\nu}\equiv\nu/\nu_0$ is the frequency relative to the central frequency of the band, $\chi_x$ is the rotation induced by patch $x$ at the central frequency, $\nu_0$, and $\phi_{x}$ is the phase change. The Stokes parameters for this simplified case are:
	\begin{eqnarray}
		\label{eq:Islit}
		& I= E_{1}E_{1}^*+E_{2}E_{2}^*=
		1+\cos(\Delta \phi)\cos(\Delta \chi) 
	\end{eqnarray}
	\begin{eqnarray}
		\label{eq:Qslit}
		Q\!=\! E_{1}E_{1}^*\!-\!E_{2}E_{2}^*\!=\!\frac{1}{2}\cos(2\chi_{\rm A})\!+\!\frac{1}{2}\cos(2\chi_{\rm B})\!+\!\cos(\Delta \phi)\cos(2 \bar{\chi})
	\end{eqnarray}
	\begin{eqnarray}
		\label{eq:Uslit}
		U\!=\! E_{1}E_{2}^*\!+\!E_{1}^*E_{2}\!=\!\frac{1}{2}\sin(2\chi_{\rm A})\!+\!\frac{1}{2}\sin(2\chi_{\rm B}) + \cos(\Delta \phi)\sin(2 \bar{\chi})
	\end{eqnarray}
	\begin{eqnarray}
		\label{eq:Vslit}
		& V=\frac{1}{i} \left[ E_{1}E_{2}^*-E_{1}^*E_{2} \right] =\sin(\Delta \phi)\sin(\Delta \chi)
	\end{eqnarray}
	where we have used the notation $\Delta \phi=\phi_{\rm A}-\phi_{\rm B}, \Delta \chi =\chi_{\rm A}-\chi_{\rm B},\bar{\chi}=(\chi_{\rm A}+\chi_{\rm B})/2$.
	In general, $\phi_{\rm A},\phi_{\rm B}$ have contributions from both the phase shift in the screen and the geometrical path-length of the two trajectories, and so $\phi_{\rm A}\neq \phi_{\rm B}$. 
	If no Faraday rotation is induced by the screen, then $\chi_{\rm A}=\chi_{\rm B}$. In such a situation we see that $V=0$ and $I^2=U^2+Q^2$. This is nothing but the standard Young's double-slit experiment in which the wave remains $100\%$ linearly polarized in its original direction and the intensity fluctuates based on $\Delta \phi$. Faraday rotation and the turbulent nature of the screen allows us to have $\chi_{\rm A}\neq \chi_{\rm B}$. Since $\Delta \chi \neq 0$, we have $V\neq 0$, meaning that a fraction of the incoming linear polarization has been converted to circularly polarized wave, depending on $\Delta \chi$. By construction, the overall polarization level in this case remains 100\%, since we have assumed the spectral band to be infinitely narrow.

	We now generalize this result to the case of an arbitrary spectral width of the detector.
	$\chi_x$ and $\phi_x$ are functions of frequency. As shown in \S \ref{sec:Polscreen}, $\phi_x$ changes randomly on the coherence bandwidth scale, $\nu_{\rm co}$. Furthermore, $\chi_x$ changes on a frequency scale $\geq \nu_{\rm co}$ (with equality for $N_{\chi}\gtrsim N_{\rm co}$, see \S \ref{sec:Polscreen}).
	The frequency scale over which the PA of a wave at $\nu_0$ changes significantly due to the average RM is, $\nu_{\rm RM}=\nu/2\bar{\chi}$ (see \S \ref{sec:meanRMPol}). Similarly the PA at $\nu_0$ changes significantly on a scale $\nu_{\rm \Delta RM}=\nu/2\Delta\chi$ ( $\nu_{\rm RM_x}=\nu/2\chi_x$) due to a rotation by $\Delta \chi$ ($\chi_x$). For various cases of interest $\Delta \chi \ll \bar{\chi},\chi_{\rm A},\chi_{\rm B}$, leading to $\nu_{\rm \Delta RM}\gg\nu_{\rm RM}, \nu_{\rm RM_A}, \nu_{\rm RM_B}$. This will affect the polarization signal, as we show below.
	The Stokes parameters are averages of the form 
	\begin{equation}
		\label{eq:intdef}
		\langle E_{i}E_{j}^*\rangle_{\nu_{\rm res}}  = \frac{1}{\nu_{\rm res}}\int_{\nu_0-0.5\nu_{\rm res}}^{\nu_0+0.5\nu_{\rm res}}E_iE_j^*d\nu.
	\end{equation}
	where $\nu_{\rm res}$ is the spectral bandwidth of the detector. It is constructive to define $\nu_{\rm min}=\min(\nu_{\rm res},\nu_{\rm co})$ and to divide the integral given by equation \ref{eq:intdef} to sub-sections of length $\nu_{\rm min}$. Within a frequency scale $\leq \nu_{\rm min}$, both the phases and PAs remain roughly constant and we get
	\begin{eqnarray}
		\label{eq:Imin}
		& I_{\rm min}=\langle E_{1}E_{1}^*+E_{2}E_{2}^*\rangle_{\nu_{\rm min}}=\\ &
		\int_{\tilde{\nu}_0-\frac{\tilde{\nu}_{\rm min}}{2}}^{\tilde{\nu}_0+\frac{\tilde{\nu}_{\rm min}}{2}}\left[1+\cos(\phi_{\rm A}-\phi_{\rm B})\cos(\frac{\Delta \chi}{\tilde{\nu}^2})\right]\frac{d\tilde{\nu}}{\tilde{\nu}_{\rm min}}\nonumber\\&
		\approx 1+\cos(\phi_{\rm A}-\phi_{\rm B})\cos(\Delta \chi)\min\left(1,\frac{\pi\nu_{\rm \Delta RM}}{\nu_{\rm min}}\right) \nonumber
	\end{eqnarray}
	\begin{eqnarray}
		\label{eq:Qmin}
		& Q_{\rm min}=\langle E_{1}E_{1}^*-E_{2}E_{2}^*\rangle_{\nu_{\rm min}}=\\ &
		\int_{\tilde{\nu}_0-\frac{\tilde{\nu}_{\rm min}}{2}}^{\tilde{\nu}_0+\frac{\tilde{\nu}_{\rm min}}{2}}\left[\frac{1}{2}\cos(\frac{2\chi_{\rm A}}{\tilde{\nu}^2})\!+\!\frac{1}{2}\cos(\frac{2\chi_{\rm B}}{\tilde{\nu}^2})\!+\!\cos(\phi_{\rm A}\!-\!\phi_{\rm B})\cos(\frac{2\bar{\chi}}{\tilde{\nu}^2})\right]\frac{d\tilde{\nu}}{\tilde{\nu}_{\rm min}}\nonumber\\&
		\approx \frac{1}{2}\big[\cos(2\chi_{\rm A})\min\left(1,\frac{\pi\nu_{\rm RM_A}}{2\nu_{\rm min}}\right)+\cos(2\chi_{\rm B})\min\left(1,\frac{\pi\nu_{\rm RM_B}}{2\nu_{\rm min}}\right)+\nonumber \\& 2\cos(\phi_{\rm A}\!-\!\phi_{\rm B})\cos(2 \bar{\chi})\min\left(1,\frac{\pi\nu_{\rm RM}}{2\nu_{\rm min}}\right)\big]\nonumber
	\end{eqnarray}
	\begin{eqnarray}
		\label{eq:Umin}
		& U_{\rm min}=\langle E_{1}E_{2}^*+E_{1}^*E_{2}\rangle_{\nu_{\rm min}}=\\ &
		\int_{\tilde{\nu}_0-\frac{\tilde{\nu}_{\rm min}}{2}}^{\tilde{\nu}_0+\frac{\tilde{\nu}_{\rm min}}{2}}\left[\frac{1}{2}\sin(\frac{2\chi_{\rm A}}{\tilde{\nu}^2})\!+\!\frac{1}{2}\sin(\frac{2\chi_{\rm B}}{\tilde{\nu}^2})\!+\!\cos(\phi_{\rm A}\!-\!\phi_{\rm B})\sin(\frac{2\bar{\chi}}{\tilde{\nu}^2})\right]\frac{d\tilde{\nu}}{\tilde{\nu}_{\rm min}}\nonumber\\&
		\approx \frac{1}{2}\big[\sin(2\chi_{\rm A})\min\left(1,\frac{\pi\nu_{\rm RM_A}}{2\nu_{\rm min}}\right)+\sin(2\chi_{\rm B})\min\left(1,\frac{\pi\nu_{\rm RM_B}}{2\nu_{\rm min}}\right)+\nonumber \\& 2\cos(\phi_{\rm A}\!-\!\phi_{\rm B})\sin(2 \bar{\chi})\min\left(1,\frac{\pi\nu_{\rm RM}}{2\nu_{\rm min}}\right)\big]\nonumber
	\end{eqnarray}
	\begin{eqnarray}
		\label{eq:Vmin}
		& V_{\rm min}=\frac{1}{i}\langle E_{1}E_{2}^*-E_{1}^*E_{2}\rangle_{\nu_{\rm min}}=\\ &
		\int_{\tilde{\nu}_0-\frac{\tilde{\nu}_{\rm min}}{2}}^{\tilde{\nu}_0+\frac{\tilde{\nu}_{\rm min}}{2}}\sin(\phi_{\rm A}\!-\!\phi_{\rm B})\sin(\frac{\Delta\chi}{\tilde{\nu}^2})\frac{d\tilde{\nu}}{\tilde{\nu}_{\rm min}}\nonumber\\&
		\approx \sin(\phi_{\rm A}\!-\!\phi_{\rm B})\sin(\Delta \chi)\min\left(1,\frac{\pi\nu_{\rm \Delta RM}}{\nu_{\rm min}}\right)\nonumber
	\end{eqnarray}
	As expected, the results converge with the Stokes parameters given by equations \ref{eq:Islit},\ref{eq:Qslit},\ref{eq:Uslit},\ref{eq:Vslit} in the limit of $\bar{\chi},\Delta \chi \ll \tilde{\nu}_{\rm min}^{-1}$ (for which $\chi_{\rm A},\chi_{\rm B}$ do not evolve with frequency within the spectral range $\nu_{\rm min}$). 
	When $\bar{\chi},\Delta \chi \gg \tilde{\nu}_{\rm min}^{-1}$, the integrands in equations \ref{eq:Qmin}, \ref{eq:Umin}, \ref{eq:Vmin} oscillate many times within the interval $\nu_{\rm min}$ and their amplitude is therefore strongly suppressed compared to that of equation \ref{eq:Imin}. This leads to a linear polarization $\Pi_{\rm lin,min}=\sqrt{Q_{\rm min}^2+U_{\rm min}^2}/I_{\rm min}\sim\nu_{\rm RM}/\nu_{\rm min}\ll 1$ and to a circular polarization $\Pi_{\rm cir,min}=|V_{\rm min}/I_{\rm min}|\sim \nu_{\rm \Delta RM}/\nu_{\rm min}\ll 1$ (where $\phi_{\rm A},\phi_{\rm B}$ have been taken to be randomly distributed in the interval [$0,2\pi$], and we omitted factors of order unity for clarity).
	
	Finally, we can extend the calculation over the entire bandwidth of the detector. This can be done by dividing the bandwidth into $N\sim \nu_{\rm res}/\nu_{\rm min}$ segments of length $\nu_{\rm min}$. Within each segment, equations \ref{eq:Imin}, \ref{eq:Qmin}, \ref{eq:Umin}, \ref{eq:Vmin} can be used to estimate the Stokes parameters. Between different intervals, the phases $\phi_{\rm A}, \phi_{\rm B}$ change by an amount on the order of $\pi$ and the PAs, $\chi_{\rm A}, \chi_{\rm B}$ change by an amount of order $\Delta \chi$. Since $I_{\rm min}$ has a term independent of $\phi_{\rm A}, \phi_{\rm B}, \chi_{\rm A}, \chi_{\rm B}$, it adds up coherently such that $I=NI_{\rm min}$. The other Stokes parameters are trigonometric functions of the phases and PAs and therefore add up through a random walk process $Q\approx \sqrt{N}Q_{\rm min}, U\approx \sqrt{N}U_{\rm min}, V\approx \sqrt{N}V_{\rm min}$. The result is
	\begin{equation}
		\label{eq:Pilintoy}
		\Pi_{\rm lin,\nu}\approx \Pi_{\rm lin,min}N^{-1/2}=\left(1+\frac{\min(\nu_{\rm co},\nu_{\rm res})}{\nu_{\rm RM}}\right)^{-1} \left(1+\frac{\nu_{\rm res}}{\nu_{\rm co}}\right)^{-1/2}
	\end{equation}
	\begin{equation}
		\label{eq:Picirtoy}
		\Pi_{\rm cir,\nu}\approx \Pi_{\rm cir,min}N^{-1/2}=\left(1+\frac{\min(\nu_{\rm co},\nu_{\rm res})}{\nu_{\rm \Delta RM}}\right)^{-1} \left(1+\frac{\nu_{\rm res}}{\nu_{\rm co}}\right)^{-1/2}
	\end{equation}
	For a turbulent screen, $\Delta \chi\approx \chi_0 (R_{\rm screen}/L)^{2\gamma+1\over 2}$, where by construction $R_{\rm screen}/L\leq 1$ and for a wide region of parameter space $R_{\rm screen}/L\ll 1$ (see \S \ref{sec:length}). The implication is that it is natural to expect $\chi_0\gg \Delta \chi$. If in addition $\nu_{\rm RM}<\min(\nu_{\rm co},\nu_{\rm res})$ (this is not trivial, see \S \ref{sec:timescales}), this leads to $\Pi_{\rm circ}\gg \Pi_{\rm lin}$ (with the ratio between them becoming as large as $(L/R_{\rm screen})^{2\gamma+1\over 2}$ if also $\nu_{\rm \Delta RM}<\min(\nu_{\rm co},\nu_{\rm res})$).
	An additional way to achieve $\Pi_{\rm cir}\gg \Pi_{\rm lin}$, which does not rely on having $\nu_{\rm RM}<\min(\nu_{\rm co},\nu_{\rm res})$ is naturally obtained when depolarization results from time averaging of the signal in the case where $t_{\rm res}\ll t_{\rm scr,var}$ as discussed in \S \ref{sec:Polscreen} below.
	
	\subsection{Polarization signature from scintillating screen}
	\label{sec:Polscreen}
	Having established the characteristic length and time scales associated with the source-screen-detector system we are now set to explore the polarization signal from a scintillating screen. As there are many different possible regimes to explore (depending on the ordering of the different scales) it is useful to consider a specific ordering, as an example, and derive polarization properties for that case. Based on the estimates in \S \ref{sec:length},\ref{sec:timescales}, a realistic such ordering is $\ell_{\phi}\ll \ell_{\chi}\ll R_{\rm sc}$, $\nu_{\rm co}\ll\nu_{\rm RM}$. We focus primarily on this ordering and later discuss implications for different orderings of these spatial scales. 
	
	For a 2D screen, we can write the electric vector at the observer location as
	\begin{equation}
		{\bf E_\nu(x_0, \rm t)} \sim \left[\frac{\ell_\phi}{R_{\rm sc}}\right] \sum_{j=1}^{N_{\phi}} e^{i \phi_{\ell_\phi}^{\rm j}}\, {\bar{\bar T}}(\chi^{\rm j})
		{\bf E_{\nu 0}(\rm t_0}),
		\label{E-pol}
	\end{equation}
	where $N_{\phi}\approx (R_{\rm sc}/\ell_{\phi})^2$ is the number of independent patches in the scattering screen from which the observer receives radiation. Note that the number of patches which rotate the wave by an independent amount is $N_{\chi}=(R_{\rm sc}/\ell_{\chi})^2\ll N_{\phi}$, so many of the $\chi^{\rm j}$s are the same (see figure \ref{fig:schem}).
	As we show below, it is, nonetheless, typically the smaller scale, $\ell_{\phi}$ which determines the polarization properties.
	
\begin{figure*}
		\centering
		\includegraphics[width = 0.4\textwidth]{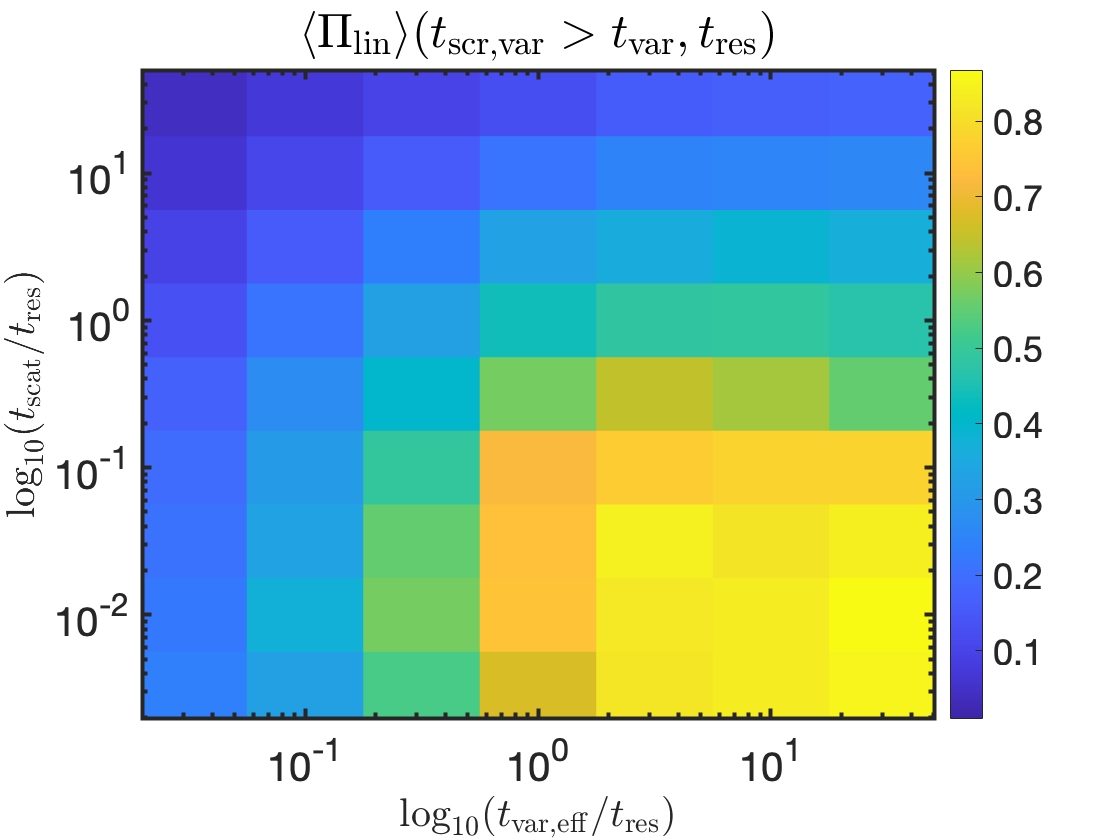}
		\includegraphics[width = 0.4\textwidth]{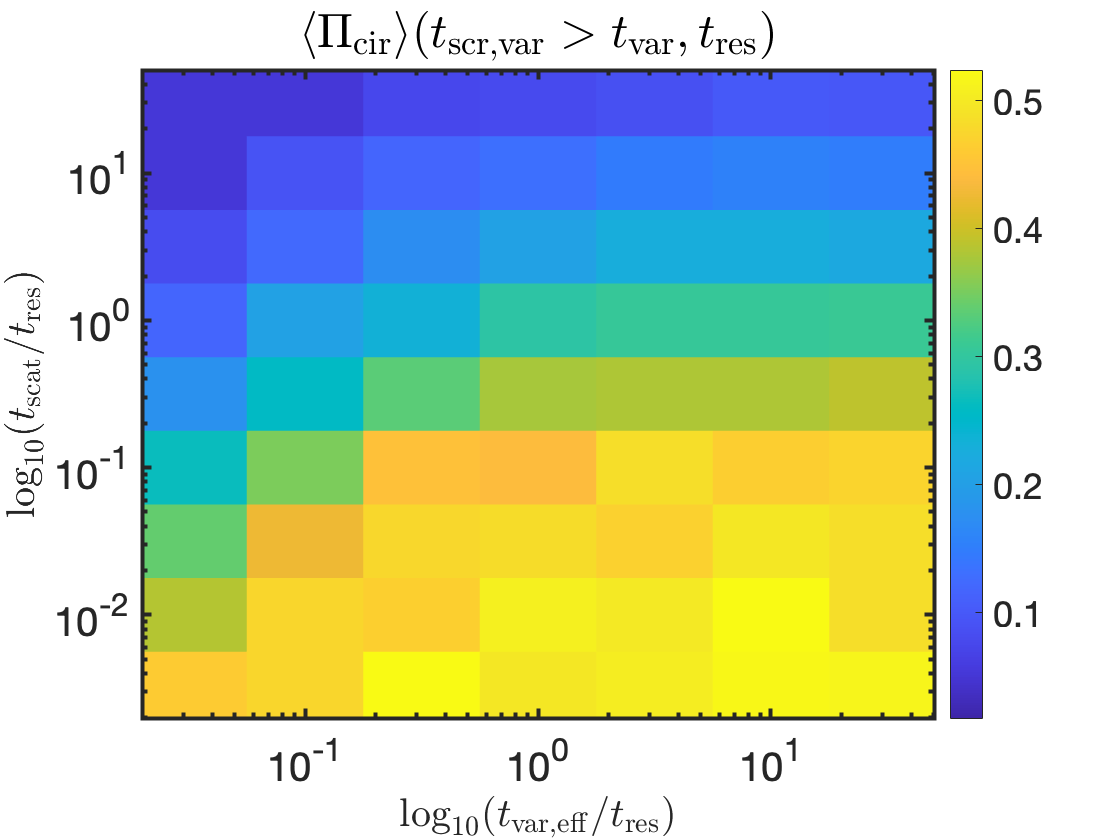}\\
		\includegraphics[width = 0.4\textwidth]{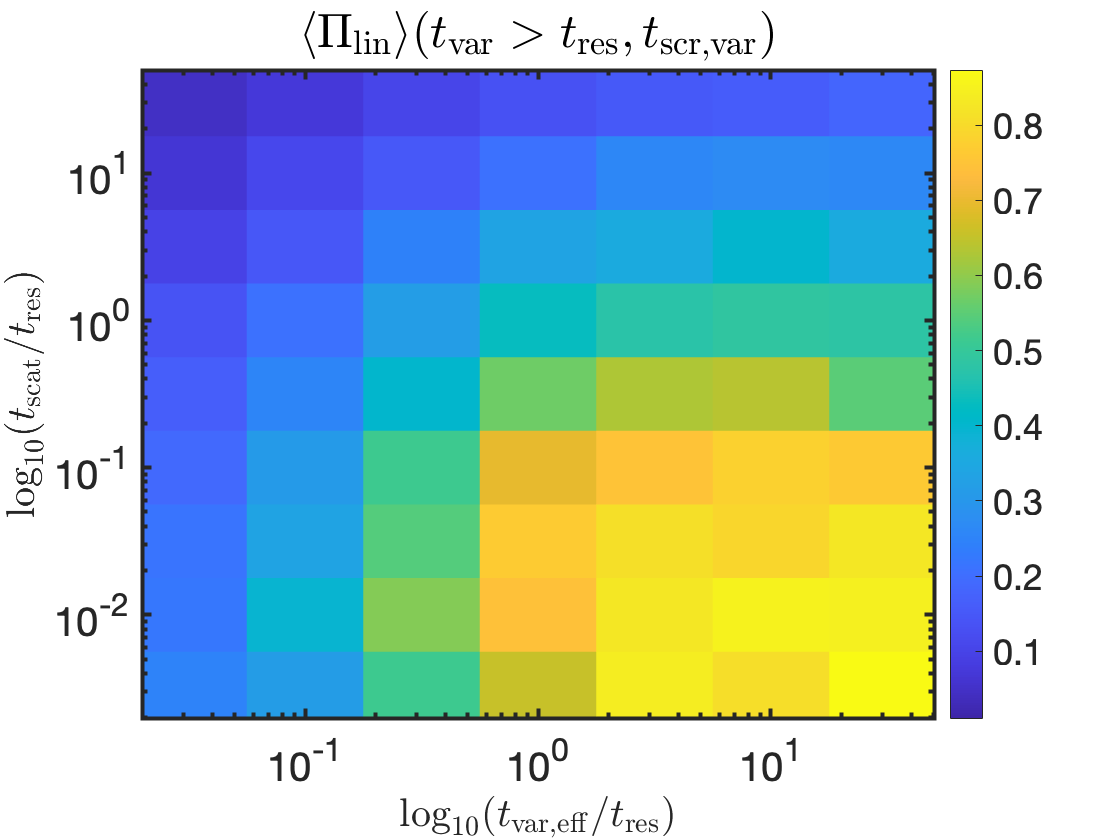}
		\includegraphics[width = 0.4\textwidth]{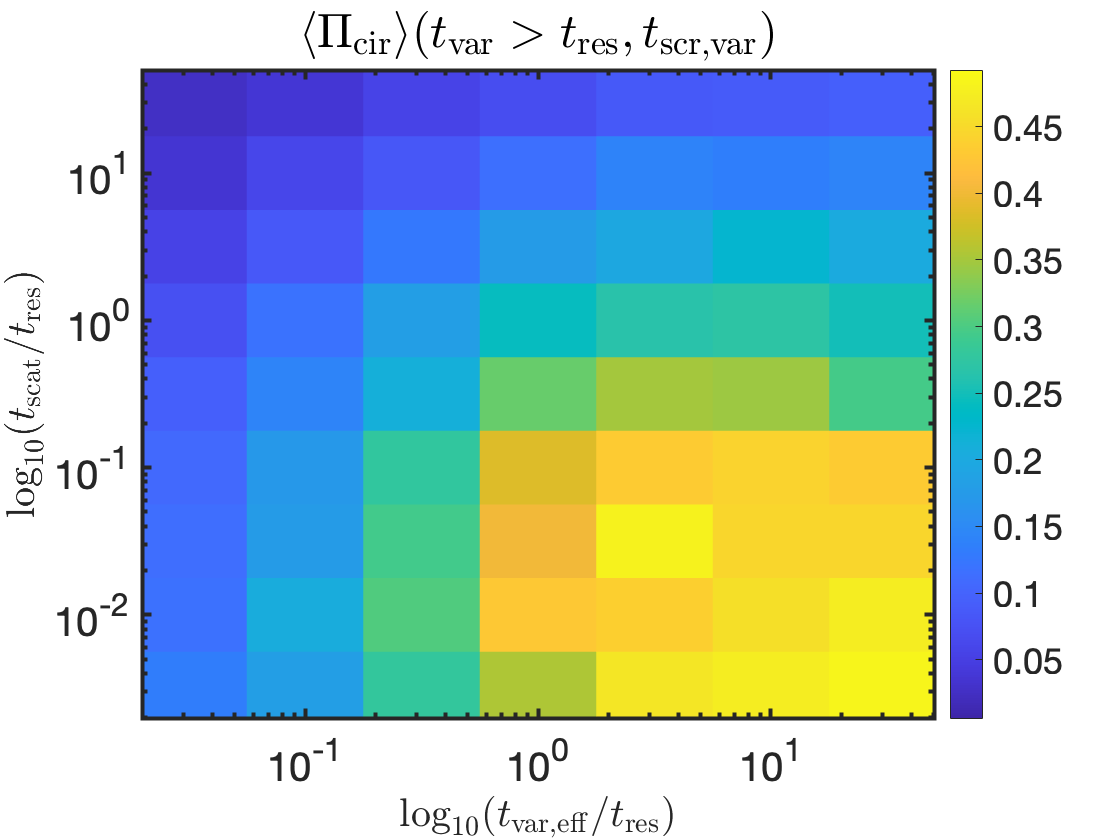}
		\caption{Results of a Monte Carlo calculation showing the degree of polarization for a temporally variable wave passing through a inhomogeneous 1D screen (with inhomegeneities each causing a rotation of the polarization vector by $\Delta \chi \sim \pi$). The linear (circular) polarization is shown in the left (right) panel. The top panels show results for the case in which the temporal variability is dominated by the source, $t_{\rm var,eff}=t_{\rm var}$ \& $t_{\rm scr,var}>t_{\rm res},t_{\rm var}$, and the bottom panel for the case when it is dictated by the screen, $t_{\rm var,eff}=t_{\rm scr,var}$ \& $t_{\rm var}>t_{\rm res},t_{\rm scr,var}$. The characteristic time delay associated with the screen is $t_{\rm scat}$ and the temporal resolution of the detector is $t_{\rm res}$. The spectral resolution of the detector is $\nu_{\rm res}=t_{\rm res}^{-1}$. }
		\label{fig:1D}
	\end{figure*}

	As described in \S \ref{sec:toymodelPi}, if the EM source is 100\% linearly polarized, the source PA is stable with time, and the observing bandwidth is infinitesimally narrow, then the wave after crossing the screen will remain 100\% polarized although the PA would have rotated and there would be, in general, some non-zero circularly polarized component \footnote{This statement assumes that the properties of the screen remain fixed in time and that the screen's position is static relative to the observer. When this is not the case, and $t_{\rm res}\gg t_{\rm scr,var}$ then, similar to the case $t_{\rm res}\gg t_{\rm var}$ analyzed in \S \ref{sec:tempvar}, the result is strong temporal depolarization of the detected wave.}. If the effect of the scattering screen is to randomly change the PA and the phase of the EM wave (in an uncorrelated fashion), then the result would be a PA pointing in a random direction on the Poincare sphere \footnote{The exact distribution of the PA pointing on the Poincare sphere is explicitly calculated numerically below.}, leading to mean square values for the linear and circular polarizations (the root of which represents typically expected values) of
	\begin{eqnarray}
		\label{eq:Poincre}
		&    \langle\Pi_{\rm lin}^2\rangle=\langle\frac{Q^2+U^2}{I^2}\rangle=\frac{\int \int \cos^2(2\chi)d\Omega}{4\pi}=\frac{2}{3}\nonumber \\ 
		&    \langle\Pi_{\rm cir}^2\rangle=\langle\frac{V^2}{I^2}\rangle=\frac{\int \int \sin^2(2\chi)d\Omega}{4\pi}=\frac{1}{3}.
	\end{eqnarray}
	The implication is that even when the screen induces no overall depolarization, it can change a linearly polarized signal to an elliptically polarized one, with a non-negligible degree of circular polarization (as demonstrated in figure \ref{fig:1D} discussed below).

	The scattering screen reduces the polarization degree when either the observing bandwidth is larger than the coherence bandwidth of the scattering screen (defined below), or the time interval over which the signal is integrated is larger than the variability time associated with temporal changes in the screen properties or its position relative to the observer (the temporal variability of the source has a much weaker affect on the total polarization degree, as discussed below). In either case, during the observing window (spectral or temporal, appropriately), the wave electric vector direction has fluctuated as the observed radiation has suffered different degrees of Faraday rotations from {\it eddies} which decohered in frequency or time. 
	
	The coherence bandwidth of the scattered wave, $\nu_{\rm co}$, is related to the scattering time $t_{\rm scat}$ via the uncertainty principle: $\nu_{\rm co} \sim t_{\rm scat}^{-1}$. In the limit of strong scintillation, $\nu_{\rm co}\ll \nu_0$ \footnote{As shown in \S \ref{sec:timescales}, $t_{\rm scat},\nu_{\rm co}$, are generally wavelength dependent, but for $\nu_{\rm res}\ll \nu_0$ they change very little across the detector's band, and therefore, we take $t_{\rm scat}= t_{\rm scat}(\nu_0), \nu_{\rm co} = \nu_{\rm co}(\nu_0)$}. Indeed, this is the reason that the finite spectral resolution can play a role in determining the polarization, despite the fact that $\nu_{\rm res}\ll \nu_0$. This can be understood by recognizing the very large geometrical phase shift between rays coming from different parts of the screen. This shift, $d\phi$, between a path along the line of sight to the center of the screen and one that intersects the screen at a distance $x$ from the center, is $d\phi = \pi x^2/(d \lambda) = \pi (x/R_{\rm F})^2$.  Taking $x\approx R_{\rm sc}$, we get $d\phi \sim \pi^{-1} (R_{\rm F}/\ell_{\phi})^2 \gg \pi$. Since $(R_{\rm F}/\ell_{\phi})^2$ is in general wavelength dependent, we see that $\nu$ needs to change by a small amount, of order $d\nu\sim \nu_0 \pi^2 (\ell_{\phi}(\nu_0)/R_{\rm F}(\nu_0))^2 \sim \nu_{\rm co}$, for the geometric phase difference between two paths, and for frequencies separated by $\nu_{\rm co}$, to become of order $\pi$.
	
	The detector has finite temporal resolution, $t_{\rm res}$, and spectral resolution $\nu_{\rm res}$. In general (as per the uncertainty principle), $t_{\rm res} \nu_{\rm res}\geq 1$. We focus below on the optimal detector for which the above relation is an equality.
	
	The intrinsic temporal variability, $t_{\rm var}$, and the temporal variability associated with the screen, $t_{\rm scr,var}$ have distinct effects on the observed polarization. On a timescale, $t_{\rm scr,var}$, the phases and PAs of wavelets from all points of the screen have significantly changed relative to their previous values. This means that within every segment of duration $t_{\rm scr,var}$, the polarization is reduced as per equations \ref{eq:Pilintoy}, \ref{eq:Picirtoy}, and the contributions from those segments add up as a random walk (see \S \ref{sec:tempvar}) such that $\Pi=\Pi_{\nu}\cdot (1+t_{\rm res}/t_{\rm scr,var})^{-1/2}$. Interestingly, the temporal variability due to the source, has a distinct effect on the observed polarization. The reason is that while the PA incident on the screen may be rapidly changing with time, the PA difference of wavelets exiting through different parts of the screen ($\Delta \chi$) is not affected by this change. Since $\Pi_{\rm cir}$ depends only on $\Delta \chi$ (see \S \ref{sec:toymodelPi}), the observed circular polarization remains unchanged in the presence of rapid source variability, i.e. it reflects the intrinsic circular polarization of the source. Note, however, that the source variability does reduce the linear polarization (in much the same way that the screen variability does). To summarize, we find that the polarization degree, assuming $\nu_{\rm co}\ll\nu_{\rm RM}$ (implying no depolarization due to the mean RM of the screen -- the opposite limit is addressed at the end of the section) is well approximated by
	\small
	\begin{eqnarray}
		\label{eq:Polgen}
		& \Pi_{\rm lin}\!\approx\! \bigg(\frac{\nu_{\rm res}}{\nu_{\rm co}}\!+
		\!1\bigg)^{-1/2}\bigg(\frac{t_{\rm res}}{t_{\rm var, eff}}\!+\!1\bigg)^{-1/2}\!=
		\!\bigg(\frac{t_{\rm scat}}{t_{\rm res}}\!+\!1\bigg)^{-1/2}\bigg(\frac{t_{\rm res}}{t_{\rm var, eff}}\!+\!1\bigg)^{-1/2} \\
		&   \Pi_{\rm cir}\!\approx\! \bigg(\frac{\nu_{\rm res}}{\nu_{\rm co}}\!+
		\!1\bigg)^{-1/2}\bigg(\frac{t_{\rm res}}{t_{\rm scr,var}}\!+\!1\bigg)^{-1/2}\!=
		\!\bigg(\frac{t_{\rm scat}}{t_{\rm res}}\!+\!1\bigg)^{-1/2}\bigg(\frac{t_{\rm res}}{t_{\rm scr,var}}\!+\!1\bigg)^{-1/2} \nonumber.
	\end{eqnarray}
	\normalsize
	where the last equality in each line holds for $\nu_{\rm res}=t_{\rm res}^{-1}$.
	This result is reproduced numerically and presented in figure \ref{fig:1D} in which we depict the observed degree of polarization for an initially linearly polarized wave passing through a static 1D screen with properties as described above \footnote{We calculate results for a 1D screen since a 2D screen is computationally much more intensive, while the underlying physics is qualitatively unchanged between a 1D and 2D screens (we mention explicitly whenever there are quantitative differences between the two).}. A significant amount of depolarization is obtained unless $t_{\rm scr,var}>t_{\rm res}$ {\it and} $t_{\rm res}> t_{\rm scat}$. As suggested by equation \ref{eq:Poincre} we find that the screen always leads to a significant fraction of the signal becoming circularly polarized. In particular, when $t_{\rm scr,var}<t_{\rm var}$, $\langle \Pi_{\rm lin}^2/\Pi^2\rangle=2/3$, $\langle \Pi_{\rm cir}^2/\Pi^2\rangle=1/3$. This holds even for $t_{\rm var, eff}>t_{\rm res}>t_{\rm scat}$ for which no overall depolarization is induced by the screen. When $t_{\rm scr,var}>t_{\rm var}$ the circular to linear polarization ratio increases as $\Pi_{\rm cir}^2/\Pi_{\rm lin}^2\approx (t_{\rm var}+t_{\rm res})t_{\rm scr,var}/[(t_{\rm scr,var}+t_{\rm res})t_{\rm var}]$ (see equation \ref {eq:Polgen}).

	It is at first glance perhaps counter-intuitive that the spectral depolarization by the screen is dictated by the coherence bandwidth associated with the patches of size $\ell_{\phi}$, despite the outgoing PA from many of these different patches being the same (due to $\ell_{\phi}\ll \ell_{\chi}$ or equivalently $N_{\phi}\gg N_{\chi}$). The reason for this is that within the observed spectral band of width $\nu_{\rm res}$ there are $N_{\rm co}\equiv \nu_{\rm res}/\nu_{\rm co}$ independent segments of frequency. Within each of those frequency segments the phase and amplitude contributions from all patches on the screen (of size $\ell_{\phi}$), and the resulting PA from their superposition, is significantly changed compared to the other frequency segments. 
	Therefore, as long as $N_{\chi}\gtrsim N_{\rm co}>1$, the $N_{\rm co}$ segments are likely to all have independent PAs and the overall depolarization will be set by random walk as $N_{\rm co}^{-1/2}$. As shown in figure \ref{fig:polarizationvsmean}, this is expected to hold for a large area of parameter space even when $\ell_{\chi}\gg \ell_{\phi}$. Alternatively, if $N_{\rm co}\gtrsim N_{\chi}>1$, the number of independent PAs in the superposition is reduced and limited by $N_{\chi}$. The signal is still depolarized, but to a lesser extent, with $\Pi\propto N_{\chi}^{-1/2}$. In the general case, the spectral depolarization is given by $[1+\min(N_{\rm co},N_{\chi})]^{-1/2}$. We define a characteristic frequency, $\nu_{\rm N}$, such that $N_{\chi}(\nu_{\rm N})=N_{\rm co}(\nu_{\rm N})$:
	\begin{eqnarray}
		\label{eq:nuN}
		\nu_{\rm N}=\left(\frac{2cd}{\mathcal{R}}\right)^{2\gamma+1\over 2\gamma+9} \left( {2\pi m^2 c^2 \over q^3 n_{\rm e} B_{||}}\right)^{-4\over 2\gamma+9}\ell_{\rm max}^{-4\gamma\over 2\gamma+9}L^{2\over 2\gamma+9}
	\end{eqnarray}
	For $\nu<\nu_{\rm N}$ the spectral depolarization is given by $(1+\nu_{\rm res}/\nu_{\rm co})^{-1/2}$ while at greater frequencies the polarization is increased relative to this value (i.e. less depolarization by the screen), due to the fact that there are an insufficient number of patches with independent PAs. Another limiting case is when there is only one PA produced by the entire screen, i.e. $\ell_{\chi}=R_{\rm sc}$. As shown in \S \ref{sec:length}, this occurs above the characteristic frequency $\nu_{\chi\rm s}$. We point out that so long as $\nu_{\rm co}<\nu_{\rm res}$ (as is necessary for spectral depolarization) it is guaranteed that $\nu_{\rm N}<\nu_{\chi\rm s}$. 
	
	When $\ell_{\chi} \gtrsim R_{\rm sc}$, waves passing through different parts of the screen visible to us emerge with nearly the same PA, i.e. their electric field vectors have rotated by nearly the same amount. Therefore, the screen causes almost no spectral depolarization. However, the induced circular polarization is non-zero, but it decreases with increasing $\ell_{\chi}/R_{\rm sc}\gtrsim 1$. This is because the non-zero RM gradient results in differential RM across the visible screen of magnitude $\Delta\chi\sim (R_{\rm sc}/\ell_{\chi})$, which leads to non-zero induced circular polarization
	\begin{equation}
		\Pi_{\rm circ} \sim R_{\rm sc}/\ell_{\chi} 
	\end{equation}
	as can be easily calculated using equation (\ref{eq:Vslit}).
	
\begin{figure}
		\centering
		\includegraphics[width = 0.4\textwidth]{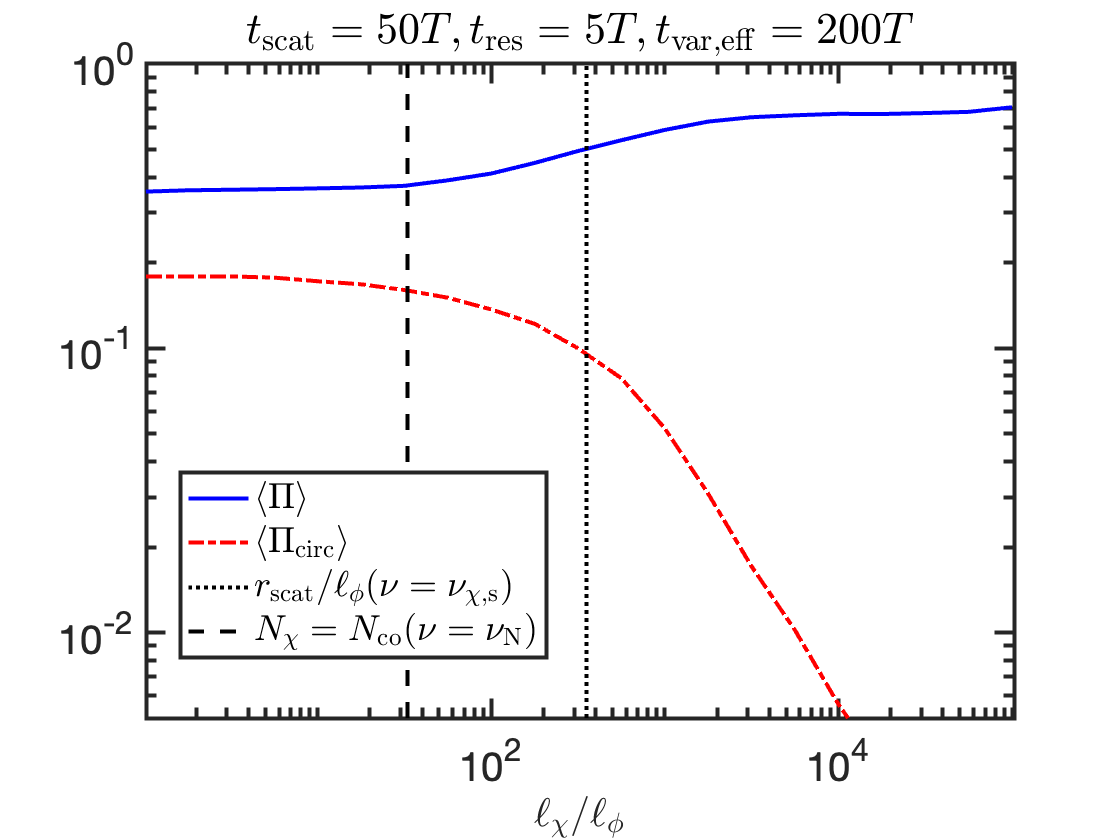}
		\includegraphics[width = 0.4\textwidth]{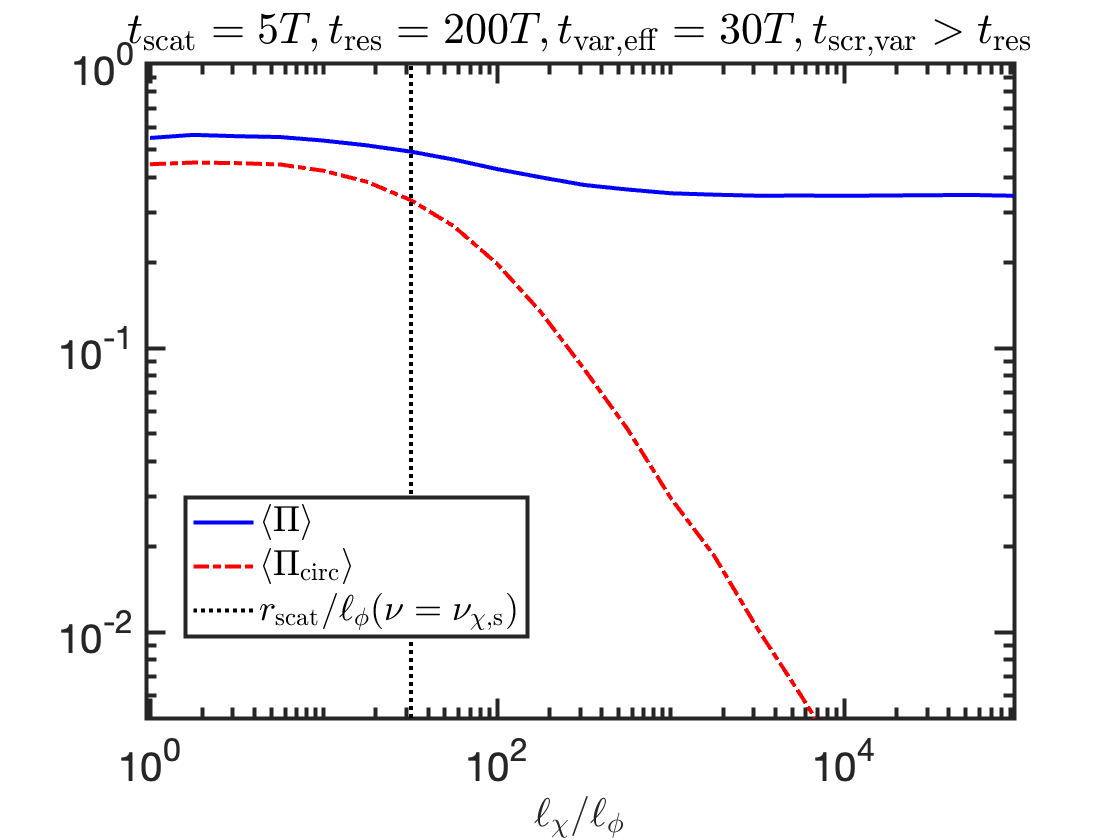}
		\caption{Results of Monte Carlo calculations showing the total degree of polarization (solid line) and degree of circular polarization (dot-dashed) for a wave passing through a 1D screen, as a function of the ratio of the typical scales for electric field rotation ($\ell_{\chi}$), and phase change ($\ell_{\phi}$). The scattering radius relative to the eddy size is shown as a dotted vertical line. The different panels show different orderings of the characteristic timescales. For $\ell_{\chi}>R_{\rm sc}$ the circular polarization falls off as $\ell_{\chi}^{-1}$ as described in \S \ref{sec:Polscreen}.}
		\label{fig:lchilpi}
	\end{figure}

	Figure \ref{fig:lchilpi} shows how the polarization degree changes with $\ell_{\chi}/\ell_{\phi}$. For $t_{\rm scat}>t_{\rm res}$ and $t_{\rm var, eff}>t_{\rm res}$, the depolarization is dominated by spectral decoherence (as in \S \ref{sec:nuvar}). As explained above, this leads to an increase in $\langle \Pi \rangle$ and a decrease in $\langle \Pi_{\rm circ} \rangle$ when $N_{\chi}<N_{\rm co}$, and the depolarization vanishes in the limit $\ell_{\chi}\gg R_{\rm sc}$. For $t_{\rm scat}<t_{\rm res}$ and $t_{\rm var, eff}<t_{\rm res}<t_{\rm scr,var}$, the depolarization is governed by the source's temporal variability. Indeed, the level of polarization for this case is the same as for the equivalent case with temporal variability only and no screen, described in \S \ref{sec:tempvar}.

	\begin{figure}
		\centering
		\includegraphics[width = 0.4\textwidth]{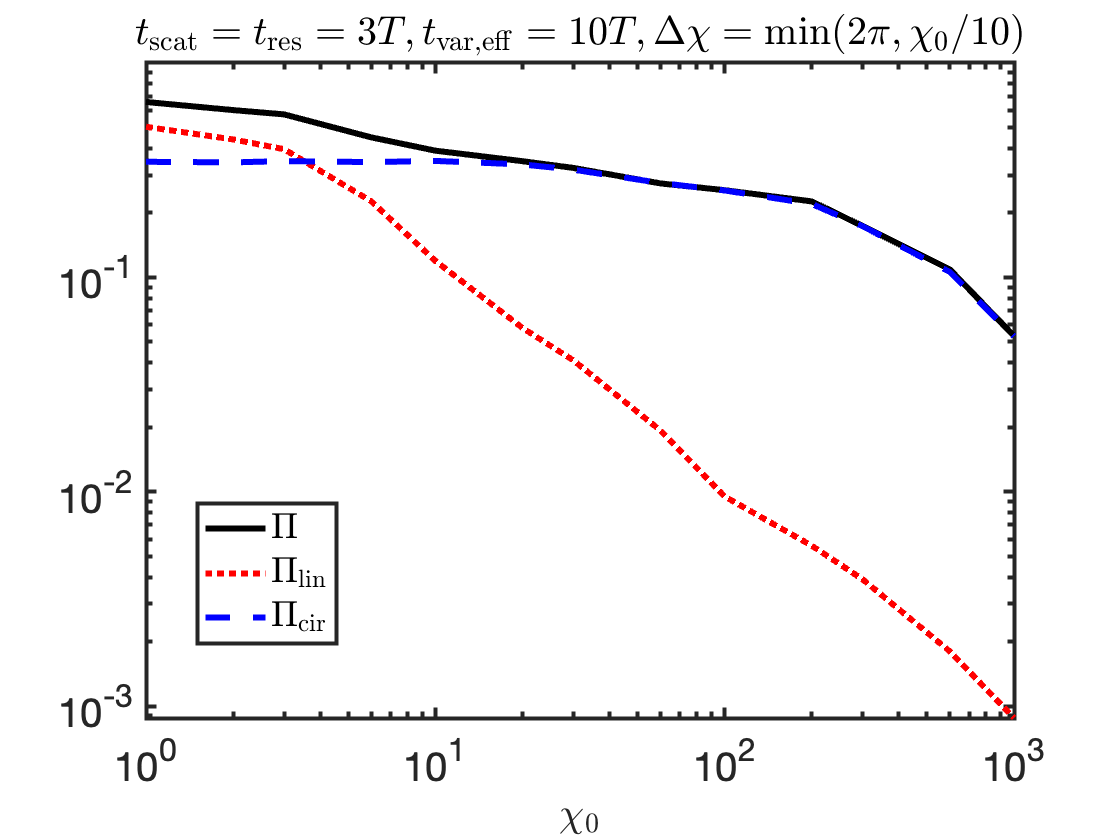}
		\caption{ Results of Monte Carlo calculations showing the total and circular degrees of polarization for a screen and wave with fixed characteristic timescales, and with varying mean values of rotation induced by the screen at the central frequency, $\chi_0$. Parameters assumed for this calculation are $t_{\rm res}=t_{\rm scat}=3T, t_{\rm var}=10T$. The typical difference in the rotation induced by two random patches in the screen separated by a distance $\gg \ell_{\chi}$, is taken to be $\Delta \chi=\min(\pi,\chi_0/10)$. At large $\chi_0,\Delta \chi$, the expected scaling $\Pi_{\rm lin}\propto \chi_0^{-1},\Pi_{\rm cir}\propto \Delta \chi^{-1}$ is recovered.}
		\label{fig:chi0dep}
	\end{figure}
	
	So far in this section we have focused on the case $\nu_{\rm co}\ll\nu_{\rm RM}$. In the opposite limit, provided that $\nu_{\rm RM}<\nu_{\rm res}$, the degree of {\it linear} polarization is reduced relative to equation \ref{eq:Polgen} by a factor of $\min(\nu_{\rm co},\nu_{\rm res})/\nu_{\rm RM}$. Interestingly, the degree of circular polarization is suppressed by a smaller factor of $\min(\nu_{\rm co},\nu_{\rm res})(R_{\rm screen}/L)^{2\gamma+1\over 2}/\nu_{\rm RM}$ (see \S \ref{sec:toymodelPi}). Since typically $R_{\rm screen}\approx R_{\rm sc}\ll L$ (see figure \ref{fig:spatialscales}), it is natural to have $\Pi_{\rm cir}\gg \Pi_{\rm lin}$ in this regime. This is demonstrated in figure \ref{fig:chi0dep}.
	
	Our discussion so far was focused on the strong scattering regime. In the weak scattering regime, it is still possible to have an effective depolarization by the screen if the mean rotation introduced by it, is $\chi_0 \gg 1$, causing significant wave rotation, as long as $\nu_{\rm RM}\ll \nu_{\rm res}$ (see \S \ref{sec:meanRMPol}). Importantly, as opposed to the sources of depolarization discussed above, the source of this depolarization is not stochastic, but rather the secular dependence of the field rotation on frequency. This depolarization can therefore be removed (and indeed this is commonly done) from the data by accounting for the RM and is in that sense qualitatively separate from the stochastic sources of depolarization discussed in this work.
	
	\subsection{Frequency dependence of polarization properties}
	\label{sec:Polsummary}
	Summarizing our results from \S \ref{sec:meanRMPol}, \ref{sec:Polscreen}, we write the degree of polarization for an incoming linearly polarized wave,
	\begin{equation}
		\label{eq:Pisummarizetot}
		\sqrt{\langle \Pi^2\rangle}=\Pi_t \Pi_{\nu}\!\approx \!\left\{ \begin{array}{ll} \left(1\!+\!\frac{t_{\rm res}}{t_{\rm scr,var}}\right)^{-{1\over 2}} \left(1\!+\!\min(N_{\rm co},N_{\chi})\right)^{-{1\over 2}}& \nu\ll \nu_{\rm \chi s}\\ \times \left(1+\frac{\min(\nu_{\rm co},\nu_{\rm res})}{\nu_{\rm RM} (L/R_{\rm screen})^{2\gamma+1\over 2}}\right)^{-1}& \\  \left(1+\frac{t_{\rm res}}{t_{\rm var}}\right)^{-{1\over 2}} \left(1+\frac{\nu_{\rm res}}{\nu_{\rm RM}}\right)^{-1} & \nu\gg \nu_{\rm \chi s}
		\end{array} \right. 
	\end{equation}
	\begin{equation}
		\label{eq:Pisummarizelin}
		\sqrt{\langle \Pi_{\rm lin}^2\rangle}\approx A_{\rm lin}\left\{ \begin{array}{ll} \left(1\!+\!\frac{t_{\rm res}}{t_{\rm var,eff}}\right)^{-{1\over 2}} \left(1\!+\!\min(N_{\rm co},N_{\chi})\right)^{-{1\over 2}}& \nu\ll \nu_{\rm \chi s}\\ \times \left(1+\frac{\min(\nu_{\rm co},\nu_{\rm res})}{\nu_{\rm RM}}\right)^{-1}& \\  \left(1+\frac{t_{\rm res}}{t_{\rm var}}\right)^{-{1\over 2}} \left(1+\frac{\nu_{\rm res}}{\nu_{\rm RM}}\right)^{-1} & \nu\gg \nu_{\rm \chi s}
		\end{array} \right. 
	\end{equation}
	\begin{equation}
		\label{eq:Picsummarizecir}
		\sqrt{\langle \Pi_{\rm cir}^2\rangle}\!\approx\! A_{\rm cir} \left\{ \begin{array}{ll} \left(1\!+\!\frac{t_{\rm res}}{t_{\rm scr,var}}\right)^{-{1\over 2}} \left(1\!+\min(N_{\rm co},N_{\chi})\right)^{-{1\over 2}} & \nu\ll \nu_{\rm \chi s}\\ \times \left(1+\frac{\min(\nu_{\rm co},\nu_{\rm res})}{\nu_{\rm RM} (L/R_{\rm screen})^{2\gamma+1\over 2}}\right)^{-1} \\  \Pi_{\rm cir}(\nu=\nu_{\chi,\rm s}) \cdot\left(\frac{R_{\rm screen}}{\ell_{\chi}}\right) & \nu\gg \nu_{\chi\rm s}
		\end{array} \right. 
	\end{equation}
	where $A_{\rm lin}=2/3$ and $A_{\rm cir}=1$ for ($\nu_{\rm RM}\ll \min(\nu_{\rm co},\nu_{\rm res})$ \& $R_{\rm screen}\ll L$) or for $t_{\rm var}\ll \min(t_{\rm scr,var},t_{\rm res})$ and $A_{\rm cir}=1/3$ otherwise.
	Note that by construction $\nu_{\rm \chi s}<\nu_*$, so strong scintillation is ensured at $\nu\ll \nu_{\rm \chi s}$. As shown in \S \ref{sec:tempvar}, for typical ISM-like screens $\nu_{\rm rs}>\nu_{\rm r,RM}$, and therefore at all frequencies $\nu_{\rm RM}>\min(\nu_{\rm co},\nu_{\rm res})$. Thus, the term in the second line of equations \ref{eq:Pisummarizetot}, \ref{eq:Pisummarizelin}, \ref{eq:Picsummarizecir} approaches unity, and consequently the polarization becomes independent of the mean RM of the screen.
	
	The frequency dependence of the observed polarization degree is given in table \ref{tbl:polfreq} for different orderings of the characteristic frequencies we have considered in this paper. We note that the observed polarization degree of an EM wave, that is 100\% linearly polarized at the source, is strongly frequency dependent after it passes through a magnetized scattering screen in most cases.
	As shown in \S \ref{sec:tempvar} and figures \ref{fig:tempscales}, \ref{fig:polarizationvsmean} both spectral and temporal depolarizations by the screen can be expected at frequencies in which FRBs are routinely observed for reasonable properties of the source-screen-detector system. The general trend is for the waves to become less polarized at lower frequencies. Finally, even when there is no depolarization caused by propagation, one must still have $\nu_{\chi \rm s}\ll\nu$ in order for the polarization to remain highly linear.

	\begin{table*}
		\centering
		\caption{\label{tbl:polfreq}
			Asymptotic scaling of the polarization degree with frequency for different orderings of the characteristic frequencies. The total polarization is given by the product of the temporal and spectral polarizations: $\Pi=\Pi_{t}\Pi_{\nu}$. We focus here on the expected case of $\nu_{\rm rs}>\nu_{\rm r,RM}$. Unless otherwise stated, $\Pi_{\rm cir}^2/\Pi_{\rm lin}^2\approx 0.5$ for the outgoing wave within a band of width $\nu_{\rm res}$. In the opposite regime, there are a large number of sub-cases, including some in which the polarization decreases with frequency over a certain interval.}
		\centering
		\resizebox{0.7\textwidth}{!}{
			\begin{threeparttable}	\begin{tabular}{ccc}\hline	
					condition & Polarization & $\alpha=\gamma=1/3$\tabularnewline
					\hline 
					$\nu<\nu_{\rm rv}, \nu<\nu_{\rm var},\nu<\nu_{\rm \chi s}$  & $\Pi_{t}\propto \nu^{2\alpha+3\over 4\alpha+2}$   & $\Pi_{t}\propto \nu^{1.1}$\tabularnewline
					$\nu<\nu_{\rm rv}, \nu>\nu_{\rm var},t_{\rm scr,var}>t_{\rm res}$  & $\Pi_{t,\rm lin}\propto \nu^{0.5},\Pi_{t,\rm cir}\approx \frac{1}{2}$ & $\Pi_{t,\rm lin}\propto \nu^{0.5}$
					\tabularnewline
					$\nu<\nu_{\rm rv}, \nu>\nu_{\rm var},t_{\rm scr,var}<t_{\rm res},\nu>\nu_{\chi,\rm s}$  & $\Pi_{t}\propto \nu^{0.5}$ & $\Pi_{t}\propto \nu^{0.5}$
					\tabularnewline
					$\nu<\nu_{\rm rv}, \nu>\nu_{\rm var},t_{\rm scr,var}<t_{\rm res},\nu<\nu_{\chi,\rm s}$  & $\Pi_{t,\rm lin}\propto \nu^{0.5},\Pi_{t,\rm cir}\propto \nu^{2\alpha+3\over 4\alpha+2}$ & $\Pi_{t,\rm lin}\propto \nu^{0.5},\Pi_{t,\rm cir}\propto \nu^{1.1}$
					\tabularnewline
					$\nu>\nu_{\rm rv}$  & $\Pi_{t}=1$  \tabularnewline
					\hline
					$\nu<\nu_{\rm rs},\nu<\nu_{\rm N}, \nu<\nu_{\rm \chi s}$  & $\Pi_{\nu}\propto \nu^{2\alpha+5\over 4\alpha+2}$ & $\Pi_{\nu}\propto\nu^{1.7}$ \tabularnewline
					$\nu<\nu_{\rm rs},\nu>\nu_{\rm N}, \nu<\nu_{\chi\rm s}$  & $\Pi_{\nu}\propto \nu^{\frac{4}{2\gamma+1}+\frac{2\alpha+3}{2\alpha+1}}$ & $\Pi_{\nu}\propto\nu^{4.6}$ \tabularnewline
					$\nu>\nu_{\chi\rm s},\nu>\nu_{\rm r,RM}$  & $\Pi_{\nu}\propto \nu^{2}$ & $\Pi_{\nu}\propto \nu^{2}$ \tabularnewline
					$\nu>\nu_{\rm rs}$ or $\nu>\nu_{\chi\rm s},\nu<\nu_{\rm r,RM}$ & $\Pi_{\nu}=1$ \tabularnewline
				\end{tabular}
			\end{threeparttable}
		}
	\end{table*}

	\section{Comparison to observations}
	\label{obs}
	
	An interesting case is that of the repeating FRB 20180301A, which was found by \cite{LuoPol}. This burst exhibited significant PA swings on a timescale of $\sim 10$\, ms while maintaining a large degree of linear polarization ($\Pi_{\rm lin}>0.7$), and having strong limits on the level of circular polarization ($\lesssim 0.03$ in some bursts). Such a behaviour is not expected for depolarization by a scattering screen (which, whenever it affects the polarization also necessarily induces a significant circular to linear polarization ratio). The PA swings for this object are very likely intrinsic properties of the source. Furthermore, the waterfall plots suggest a possible decorrelation bandwidth of $\sim 5$MHz, which is slightly larger than the spectral resolution ($\approx 2$\,MHz). This too points to depolarization by a scattering screen not being significant. Furthermore, the timescale of PA variations caused by the transverse velocity of the screen or turbulent motion of eddies within it is
	\begin{equation}
		\label{eq:tscrvarobs}
		t_{\rm scr,var}=\frac{\ell_{\phi}}{v_{\rm max}}= ({\rm 500\, s})\, \nu_{\rm co,5MHz}^{1/2}\nu_{\rm 1.25GHz}^{-1} d_{\rm kpc}^{1/2} v_{\rm max,7}^{-1}
	\end{equation}
	where $\nu_{\rm co,5MHz}\equiv \nu/\mbox{5MHz},\nu_{\rm 1.25GHz}\equiv\nu/1.25\mbox{GHz},d_{\rm kpc}=d/\mbox{kpc},v_{\rm max,7}=v_{\rm max}/10^7\mbox{cm s}^{-1}$. Equation \ref{eq:tscrvarobs} shows that for any reasonable distance of the screen, the associated variability time is likely to be several orders of magnitude greater than $10$\,ms.
	
	Many other FRBs, such as 20171209A, 20190102C, 20190608B and 20190711A, have strong upper limits on their degree of circular polarization, while showing significant degrees of linear polarization. In all these cases we must conclude that $\nu_{\chi \rm s}\ll \nu$. Other FRBs, such as 20140514A, 20160102A, 20180309A show reduced linear polarization and a non-negligible amount of circular polarization, and could be the result of a magnetized scattering screen.
	
	A necessary (but not sufficient) condition in order to have $\nu<\nu_{\chi \rm s}$ and significant circular polarization induced by the screen is that $\ell_{\chi}<\ell_{\rm max}$. As shown in equation \ref{eq:chi0} this leads to the mean rotation induced by the screen being $\chi_0>1$ or equivalently
	\begin{equation}
		\label{eq:RMlim}
		\mbox{RM}_{\rm s}\gtrsim 10 \left(\frac{\nu}{1\mbox{GHz}}\right)^2\mbox{rad m}^{-2}
	\end{equation}
	where $\nu$ is the central observed frequency. Only FRBs with an RM greater than that given by equation \ref{eq:RMlim} can have circular polarization induced by the screen at an observed frequency $\nu$. The measured $\mbox{RM}_{\rm s}$ in FRB 20160102A is consistent with this limit, while FRBs 20140514A and 20180309A have no reported $\mbox{RM}$ values.
	
	Another FRB with intriguing polarization properties is FRB 20201124A. \cite{Hilmarsson2021} have reported on observations of 20 bursts from 20201124A with the Effelsberg radio telescope. Several of the bursts show significant ($0.06-0.21$) degrees of circular polarization. Very recently, \citep{Xu2021} have reported FAST observations of the same burst, which reaffirm the high degrees of circular polarization and find that some bursts have circular polarization as large as $0.75$. \cite{Hilmarsson2021} have shown that these fractions cannot be explained by generalized Faraday conversion. Interestingly, we find here that the observed circular polarization (but not the PA swings) may be attributed to multi-path propagation through a scattering screen. First, we note that FRB 20201124A has a sufficiently large RM$\approx 605\mbox{pc cm}^{-3}$ such that the necessary condition given by equation \ref{eq:RMlim} is easily satisfied. Furthermore, the reported scattering time corresponds to a coherence bandwidth at the central frequency, $1.36$GHz, of $\nu_{\rm co}\approx 1.2$MHz, which is slightly smaller than the spectral resolution of $\nu_{\rm res}=2.5$MHz. This suggests that a slight spectral depolarization of the wave by the screen is possible and that $\Pi_{\rm cir}\lesssim 0.5(\nu_{\rm co}/\nu_{\rm res})^{1/2}\approx 0.3$. This is consistent with the measured levels of circular polarization. Finally, the bursts of FRB 20201124A with the highest degrees of circular polarization are also those with the largest RM values. This too is qualitatively consistent with the expectations for a scattering screen, as higher RM makes it easier to satisfy the required criterion $\nu<\nu_{\rm \chi s}$. We stress, however, that the PA changes seen in 20201124A cannot naturally be attributed to the scattering screen. The reason is that in at least one case the PA changes by tens of degrees over 30\,ms (between bursts 10 and 11). As shown in equation \ref{eq:tscrvarobs}, given the central observed frequency and measured coherence bandwidth, this is much too short a timescale to be attributed to a scattering screen.
	
	FRB 20121102A provides us with another interesting case. This burst has the largest RM value measured to date ($\mbox{RM}\approx 1.5\times 10^5\mbox{rad m}^{-2}$, decreasing by about $\sim 10\%$ in slightly less than a year; \citealt{Michilli+18}) and yet it exhibits close to 100\% linear polarization with the circular polarization being $\lesssim 2\%$ \citep{Michilli+18}. FRB 20121102A easily satisfies the condition given by equation \ref{eq:RMlim} for a magnetized scattering screen to convert some fraction of linear polarization at 1 GHz to circular polarization. The question is, why didn't 20121102A show some degree of circular polarization? The simplest answer might be that the large RM is due to a non-Kolmogoroff plasma in the close vicinity of the source which has effectively $\ell_\chi\gg R_{\rm sc}$. Is it possible, though, that the lack of observed circular polarization is consistent with a turbulent magneto-plasma screen with the large RM? To examine this possibility, we calculate the mean magnetic field in the screen from its RM and DM values assuming the former to be dominated by the screen
	\begin{equation}
		\label{eq:Bparallel}
		B_{\parallel}=\frac{\mbox{RM}_{\rm s}2\pi m^2 c^4}{q^3 \mbox{DM}_{\rm s}}=0.01 \left(\frac{10\mbox{pc cm}^{-3}}{\mbox{DM}_{\rm s}}\right) G. 
	\end{equation}
	Since RM$_{\rm s}$ \& DM$_{\rm s}$ are both coming from the same physical region, one might expect that they suffer the same fractional change over time, i.e. $\Delta \mbox{DM}_{\rm s}/\mbox{DM}_{\rm s}\sim \Delta \mbox{RM}_{\rm s}/\mbox{RM}_{\rm s}$ \citep{Katz2021}. Observations place an upper limit to $\Delta$DM for 20121102A of $\sim 2\mbox{pc cm}^{-3}$ during the period when RM$_{\rm s}$ changed by $10\%$. This suggests that $\mbox{DM}_{\rm s}<20\mbox{pc cm}^{-3}$. Combining this with equation \ref{eq:Bparallel} gives a lower limit on $B_{\parallel}$ in this scenario. The low degree of circular polarization is consistent with the assumption that the measured RM is due to a turbulent screen, as long as $\ell_{\chi}\gtrsim R_{\rm sc}$. To be more precise, recalling that in the limit $\ell_{\chi}>R_{\rm sc}$ we have $\Pi_{\rm circ}\approx 0.5 R_{\rm sc}/\ell_{\chi}$, a limit on the circular polarization $\Pi_{\rm circ}<X$ requires $\ell_{\chi}>2XR_{\rm sc}$. Using equations \ref{eq:lchi}, \ref{eq:Rsc} (and taking $\alpha=\gamma=1/3$) this translates to $\nu>\nu_{\chi,\rm s}(2X)^{-5/23}$. We see that the dependence on $X$ is very weak. For $X=0.02$ (as relevant for FRB 20121102A) it is sufficient to have $\nu>2\nu_{\chi,\rm s}$ to suppress the circular polarization to below the observed limits. Under the same assumptions this condition can be written as
	\begin{eqnarray}
		&  \frac{\nu_{\chi,\rm s}}{\nu}=0.26\nu_{9}^{-1}\mbox{DM}_1^{6\over 23}\mbox{RM}_5^{6\over 23}L_{\rm pc}^{-{5\over 23}}\left(\frac{d}{L}\right)^{5\over 23}\left(\frac{L}{\ell_{\rm max}}\right)^{4\over 23}<0.5 \rightarrow \nonumber \\
		& L>5\times 10^{-2} \nu_9^{-{23\over 5}}\mbox{DM}_1^{6\over 5}\mbox{RM}_5^{6\over 5}\left(\frac{d}{L}\right)\left(\frac{L}{\ell_{\rm max}}\right)^{4\over 5} \mbox{ pc}
	\end{eqnarray}
	where DM$_1\equiv \mbox{DM}_{\rm s}/10\mbox{pc cm}^{-3}$, RM$_5\equiv \mbox{RM}_{\rm s}/10^5\mbox{rad m}^{-2}$. We see that this conditions is easily satisfied for ISM-type screens at an observed frequency of 1GHz. However, because of the strong frequency dependence of the circular polarization induced by a turbulent scattering screen, if observations at lower frequencies do not reveal some level of circular polarization, the constraints can become much more severe. Ultimately, one may be able to use this argument to rule out the possibility that the large observed RM for FRB 20121102A is due to a scattering screen.

	As a final example, we consider the case of FRB 20180916B. High degrees of linear polarization, $\Pi_{\rm lin}\approx 1$, have been measured from this burst at $300-1700$\,Mhz frequency range \citep{CHIME2018,Chawla2020,Nimmo2021}. It was therefore a surprise, when \cite{Pleunis2021} recently reported a reduced degree of linear polarization, $\sim 0.3-0.7$ at $\sim 150$\,MHz. The measured scattering time of $\sim 20$\,ms at the same frequency, corresponds to a narrow coherence bandwidth, $\nu_{\rm co}\sim 20$\,Hz, well below the spectral resolution. As such, the observed reduced polarization levels are consistent with expectations from spectral depolarization due to multi-path propagation (if in addition $\nu<\nu_{\chi\rm s}$). However the lack of measured circular polarization is in tension with this possibility. A different option, is that the bursts measured at 150\,MHz suffered temporal depolarization due to intrinsic PA swings in the source. This could lead to a reduction of linear polarization with no induced circular polarization (so long as $\nu>\nu_{\chi\rm s}$). We therefore posit that the lower level of linear polarization at 150 MHz for this burst is an intrinsic property of the source.
	
	\section{Discussion and Conclusions}
	\label{sec:discuss}
	Radio sources with a high degree of intrinsic polarization might appear to have smaller polarization degree if their RM is very large such that the electric field direction rotates by order unit radian within each frequency channel of the detector. This is a trivial situation for how the observed polarization can be smaller than the intrinsic value. It can be easily checked and corrected for as it has a known scaling with frequency (see equation \ref{eq:polRM} as well as \citealt{VanStraten2002,Mckinven2021}). Also, data analysis can de-RM a source with large RM, easily, if either (i) the intra-channel rotation of the electric field direction is small or (ii) by using higher frequency data: since the PA change is $\chi_0\propto \nu^{-2}$, the intra-channel rotation of the electric field direction eventually becomes sufficiently small as the frequency increases; Once the RM has been estimated it can be used to analyze the lower frequency data. The scattering of the wave by an intervening magnetized plasma screen that is studied in this work, leads to a qualitatively and quantitatively different level of observed polarization, that may be greater or smaller than that of a wave passing through a column with the same mean RM as the screen (see e.g. figure \ref{fig:polarizationvsmean}) and that, for a large region of parameter space, is independent of the mean RM of the screen (see \S \ref{sec:Polsummary}). 
	
	We have shown that the degree of spectral depolarization by a magnetized scattering screen between the source and the observer is determined by the coherence bandwidth of the screen, $\nu_{\rm co}$, which is the inverse of the scattering time $R_F^2/(2\pi^2\nu\ell_\phi^2)$; where $R_F$ is the Fresnel scale, and $\ell_\phi$ is the smallest scale of eddies in the scattering medium which lead to an accumulated phase-change through the screen of $\sim 1$ radian. When the smallest scale of eddies that lead to accumulated rotation of the wave electric vector through the screen by $\sim 1$ radian ($\ell_\chi$) is much smaller than the size of the screen visible to the observer ($R_{\rm sc}$), the observed degree of polarization due to spectral depolarization is $\Pi_{\nu}=(1+ \nu_{\rm res}/\nu_{\rm co})^{-1/2}$. The reason for this dependence on $\nu_{\rm co}$ or $\ell_\phi$ as opposed to $\ell_\chi$ is that the electric field vector of the scattered wave at the observer switches direction randomly for two frequencies separated by $>\nu_{\rm co}$. This is due to the superposition of scattered waves from many different patches on the screen, which have random field orientations and phases that change by an angle of order one radian when $\delta\nu \gtrsim \nu_{\rm co}$ (see \S \ref{sec:toymodelPi}).
	
	An additional, `temporal' depolarization may occur if the temporal resolution of the detector is poor compared to the effective variability time, $t_{\rm var,eff}=\min(t_{\rm var},t_{\rm scr,var})$; where $t_{\rm var}$ is the intrinsic variability time of the source, and $t_{\rm scr,var}$ is the variability caused by time dependent eddies in the screen or the motion of the screen relative to the source/observer. Remarkably, there is a stark difference between the case in which the temporal variability is intrinsic to the source and the case in which it is dominated by the scattering screen. In the most physically likely scenario when $t_{\rm var}< t_{\rm scr,var}$, the linear polarization decreases as $\Pi_{t\rm,lin}=(1+ t_{\rm res}/t_{\rm var})^{-1/2}$, while the circular polarization is much larger $\Pi_{t\rm,cir}=(1+ t_{\rm res}/t_{\rm scr,var})^{-1/2}$. In the opposite regime of $t_{\rm scr,var}<t_{\rm var}$, $\Pi_{t\rm,lin}\approx 2\Pi_{t\rm,cir}\approx (1+ t_{\rm res}/t_{\rm scr,var})^{-1/2}$.
	
	The polarization signatures due to multi-path propagation discussed in this work result in a scrambling of the Stokes parameters on the Poincare sphere and can lead to a large (order unity, and in some cases much higher) ratio of circular / linear polarization (see discussion in the previous paragraph as well as in \S \ref{sec:toymodelPi}). This mechanism is markedly different from the `generalized Faraday conversion' that requires a relatively large magnetic field strength in the magneto-ionic medium through which the wave passes (and possibly additional conditions such as a relativistic population of electrons or a special field geometry) for the conversion of a linearly polarized EM wave to an observable level of circular polarization (see \S \ref{sec:Intro}). Fortunately, the two scenarios can be distinguished by their different frequency dependencies of the circular polarization.  Furthermore,  multi-path propagation through a scintillating screen causes the PA of the linearly polarized component and the orientation of the circular component, to flip stochastically on a timescale, $t_{\rm scr,var}$.

	In most cases, the frequency dependence of polarization due to multi-path propagation is very strong (see figure \ref{fig:polarizationvsmean} and table \ref{tbl:polfreq}). Therefore, broad band observations would be very useful for deciding whether less than 100\% linear polarization and / or non negligible circular polarization that have been observed for several FRBs, might be due to the scattering of waves by a {\it turbulent} magneto-ionic medium between the source and us. In general, it would require a high degree of fine tuning for a large fraction of FRBs observed at different frequencies to exhibit order unity depolarization, i.e. have polarization degrees at the level of tens of percent; such a situation -- $0\ll\Pi\ll 1$ for many bursts -- would suggest that the polarization is set at the source. A minimal requirement for a scattering screen to cause depolarization and induce circular polarization (due to multi-path propagation) is that it should have large mean rotation measure, RM$_{\rm s}>10\nu_{\rm Ghz}^{2}\mbox{rad m}^{-2}$, showing also the presence of circular polarization is more prominent at lower frequencies. In particular, circular polarization by this mechanism requires that the observed frequency correspond to the strong scintillation regime. An FRB with significant circular polarization that does not satisfy these conditions must have a different origin for the observed circular polarization \citep[e.g.,][]{LP1998}. Finally, we point out that there are at least a couple of FRBs such as 20180301A and 20201124A that exhibit PA swings on a timescale much too short to be explained by a scintillating screen. These PA swings, therefore, likely originate at the FRB source itself and provide clue to the emission physics.

	Magnetized scattering screens undoubtedly exist between us and radio sources. However, deciphering the physical properties of those screens is a non-trivial task as there are a limited number of observables available to us even in the best case scenario (e.g. dispersion measure, scattering measure, rotation measure, angular broadening), and a larger number of unknown parameters (distance to the screen, its size, its density, its magnetic field strength, the velocity of the screen or eddies within it and the minimal and maximal scale over which the turbulent cascade persists). The situation is further complicated by the fact that there could be multiple screens between us and a source with different rotation and dispersion measures such that RM might be dominated by one screen and DM by another. The present study provides tools for understanding the induced polarization due to passage of a radio wave through such screens, and should help us determine whether the observed polarization is intrinsic to the source or due to propagation through a scattering screen, and can be used to characterize the intervening screens in the latter case. This is not only desirable in its own right, but could also be instrumental for using FRBs as precise cosmological probes (e.g. \citealt{Caleb2019,Macquart20,Beniamini+21}).

	Finally, the results developed in this paper will also be relevant to astrophysical systems, other than FRBs, such as radio pulsars. The large sample of Galactic pulsars, means they can provide a wealth of high quality data. Indeed, the recent ultra wide-band observations of pulsars could be used to check for frequency evolution of the polarization properties \citep{Oswald2020}, and in particular, to check for the imprints (or lack) of multi-path propagation. 
	
	\section*{Acknowledgements}
	PB was supported by the Gordon and Betty Moore Foundation, Grant GBMF5076. This work has been funded in part by an NSF grant AST-2009619.
	
	\section*{Data Availability}
	The data produced in this study will be shared on reasonable request to the authors.

\bsp	
\label{lastpage}
\end{document}